\documentclass[article,journal]{IEEEtran}
\usepackage{amsmath,amsfonts}
\usepackage{array}
\usepackage{textcomp}
\usepackage{stfloats}
\usepackage{url}
\usepackage{verbatim}
\usepackage{graphicx}
\usepackage{multirow}
\def\BibTeX{{\rm B\kern-.05em{\sc i\kern-.025em b}\kern-.08em
    T\kern-.1667em\lower.7ex\hbox{E}\kern-.125emX}}
\usepackage{balance}
\usepackage[linesnumbered,ruled,vlined]{algorithm2e}
\usepackage{subcaption}
\usepackage{cite}
\usepackage{tikz}
\usepackage{tkz-tab}
\usepackage{pgfplots}
\usepackage{pgfplotstable}
\usepackage{caption}
\usepackage{amsmath}
\usepackage{longtable}
\usepackage{amssymb}
\usepackage{multicol}
\usepackage{anyfontsize}
\usepackage{xcolor}
\usepackage{tabularx}

\usetikzlibrary{shapes,snakes}
\usetikzlibrary{positioning}
\usepgfplotslibrary{statistics}

\usepackage{amsthm}
\newtheorem{definition}{Definition}

\usetikzlibrary{shapes,snakes}
\usetikzlibrary{positioning}
\usepgfplotslibrary{statistics}
\usetikzlibrary{calc}
\usepackage{mwe}

\begin{document}
\title{Neighbor-embedded Graph Neural Network-based Crowd Delivery Traffic Management in Smart City}

\author{Kishu Gupta, \IEEEmembership{Member, IEEE,} Deepika Saxena, \IEEEmembership{Senior member, IEEE,} Ashutosh Kumar Singh, \IEEEmembership{Senior member, IEEE}, and Chung-Nan Lee  \IEEEmembership{Member, IEEE}
	\thanks{Manuscript received 31 October 2025; accepted 14 January 2026. Date of publication 16
		March 2026; date of current version 29 May 2026. This work was supported by the National Science and Technology Council Taiwan under Grant 113-2811-E-110-014. (Corresponding author: Deepika Saxena.)}
	\thanks{Kishu Gupta is with the Department of Computer Science and Engineering, National Sun Yat-sen University, Kaohsiung, 80424, Taiwan, and also with Department of Computer Science, VIZJA University, Warsaw, 01-043, Poland. (e-mail: kishuguptares@gmail.com).}
	\thanks{Deepika Saxena is with the Department of Computer Science and Engineering, University of Aizu, Aizuwakamatsu, 9650006, Fukushima Prefecture, Japan, and also with Department of Computer Science, VIZJA University, Warsaw, 01-043, Poland. (e-mail: deepika@u-aizu.ac.jp).}
	\thanks{Ashutosh Kumar Singh is with the Department of Computer Science and Engineering, Indian Institute of Information Technology Bhopal, 462003, India, and also with Department of Computer Science, VIZJA University, Warsaw, 01-043, Poland. (e-mail: ashutosh@iiitbhopal.ac.in).}
	\thanks{Chung-Nan Lee is with the Department of Computer Science and Engineering, National Sun Yat-sen University, Kaohsiung, 80424, Taiwan. (e-mail: cnlee@mail.cse.nsysu.edu.tw).}
	\thanks{Recommended for acceptance by H. Cai.
		\\ Digital Object Identifier 10.1109/TETCI.2026.3670691}}
	
	\markboth{IEEE TRANSACTIONS ON EMERGING TOPICS IN COMPUTATIONAL INTELLIGENCE, VOL. 10, NO. 3, JUNE 2026}%
	{Shell \MakeLowercase{\textit{Gupta et al.}}: A Sample Article Using IEEEtran.cls for IEEE Journals}
	
\makeatletter
\newcommand{\removelatexerror}{\let\@latex@error\@gobble}
\def\ps@IEEEtitlepagestyle{%
	\def\@oddfoot{\mycopyrightnotice}%
	\def\@oddhead{\hbox{}\@IEEEheaderstyle\leftmark\hfil\thepage}\relax
	\def\@evenhead{\@IEEEheaderstyle\thepage\hfil\leftmark\hbox{}}\relax
	\def\@evenfoot{}%
}

\def\mycopyrightnotice{%
	\begin{minipage}{\textwidth}
		\centering \scriptsize
		2471-285X © 2026 IEEE. All rights reserved, including rights for text and data mining, and training of artificial intelligence and similar technologies.\\
		Personal use is permitted, but republication/redistribution requires IEEE permission. See https://www.ieee.org/publications/rights/index.html for more information.\\
		This article has been published in IEEE TRANSACTIONS ON EMERGING TOPICS IN COMPUTATIONAL INTELLIGENC © 2026 IEEE.
	\end{minipage}
}
\makeatother	
\maketitle

\begin{abstract}
The significant upsurge in vehicle traffic presents a considerable challenge in the pursuit of smart mobilization and transportation (SMT) worldwide. Current approaches primarily focus on vehicular traffic management through congestion prediction but fall short in addressing essential objectives such as traffic reduction and appropriate vehicle selection to alleviate congestion in smart cities ($SmCt$). To address these concerns, this work introduces a novel \textit{Neighbor-Embedded Graph Neural Network-based Crowd Delivery Traffic Management} (NeCDM) Model, comprising two key components: the Traffic Congestion Prediction Unit (TCPu) and the Traffic Observation and Management Unit (TOMu). The TCPu utilizes Graph Neural Network (GNN) optimization to accurately predict traffic flow levels at various delivery stations within $SmCt$ ecosystems. Additionally, the TOMu facilitates the intelligent selection of the most suitable delivery vehicles for fulfilling crowd delivery requests ($CDR$). This work emphasizes the potential of crowd delivery as a feasible solution for achieving SMT goals while adhering to smart city parameters ($\mathcal{SCP}$s), such as reduced carbon emissions, shorter travel times, and minimized travel distances. The proposed model achieves notable improvements in computational efficiency, including reductions of up to 4.03\% in L1 loss ($\pounds$), 16.66\% in L2 loss ($\pounds_{rmse}$), and 7.64\% in computation time.
\end{abstract}

\begin{IEEEkeywords}
Crowd delivery, delivery vehicle, neighbor-embedded graph neural network, smart city, traffic congestion control. 
\end{IEEEkeywords}

\section{Introduction} \label{sec:int}
\IEEEPARstart{T}{he} rapid increase in the number of on-road vehicles has placed significant strain on smart city ($SmCt$) traffic management systems. Recent global traffic congestion ranking reports underscore the urgent need for Smart Mobility and Transportation ($SMT$) frameworks by highlighting evolving traffic patterns and peak congestion periods across major metropolitan areas \cite{report}. $SMT$ encompasses intelligent traffic and vehicle management strategies leveraging smart resources such as satellite navigation, adaptive traffic signals, automated number plate recognition, and traffic speed cameras, to achieve high mobility while minimizing sustainability-critical parameters ($\mathcal{SCP}$) such as carbon emissions, travel distance, duration, and cost \cite{ITSC-SDN-10283852}. Crowd delivery (i.e., consolidating multiple proximate deliveries into a single route) represents a promising approach to alleviating congestion within $SmCt$ environments by reducing the total number of active vehicles. Fewer on-road vehicles not only mitigate congestion but also contribute to improved $\mathcal{SCP}$ outcomes \cite{TVT1-9392372,ITSC2-9772343,ECI1-8851407}. Additionally, heterogeneous fleets comprising vehicles with different fuel types can further enhance traffic efficiency. For example, electric vehicles (EVs) offer substantial environmental benefits through near-zero emissions but are limited by factors such as travel range, load capacity, and sensitivity to terrain and weather conditions \cite{TCSS2-9772054,TVT2-9508198,ECI2-9460321}. Selecting an appropriate delivery vehicle thus involves a complex interplay of interdependent parameters including delivery distance, the number of stations in a crowd delivery route ($CDR$), and the route type (urban/local versus peripheral/highway) since vehicle performance varies with {fuel efficiency, payload, endurance, etc.} under different operating conditions. This underscores the significance of intelligent vehicle selection in ensuring effective $SMT$ and optimizing $\mathcal{SCP}$ \cite{TVT3-9804196,IQ-HDM-TASE-10681488,ITSC3-8658122}. Consequently, integrating optimized crowd delivery strategies with suitable vehicle selection mechanisms offers substantial potential for enhanced traffic management in $SmCt$ systems.
\subsection{Observed Limitations}
In the context of $SmCt$ environments that support crowd delivery, several inherent limitations constrain the effectiveness of traffic management under $SMT$ objectives. First, existing approaches struggle to extract meaningful congestion patterns from vast, volatile, and continuously growing traffic data streams with the necessary speed and precision, thereby affecting overall system robustness. Moreover, as urban traffic networks become increasingly complex and dynamic, current models frequently fail to capture the intricate interdependencies among various $SmCt$ subsystems such as vehicles, routes, and delivery operations, which are crucial for accurate congestion assessment and management. Furthermore, most prevailing traffic management frameworks emphasize either congestion prediction or time-based optimization aspects, such as shortest travel paths or minimal delivery times, to achieve $SMT$ \cite{TVT2-9508198}. This narrow focus leads to insufficient consideration of integrated objectives such as \textit{crowd delivery optimization}, \textit{reduction of on-road delivery vehicles} through predictive analytics, and \textit{intelligent vehicle selection} based on real-time traffic conditions. Ultimately, existing methods often fall short of meeting the increasing demands for higher accuracy, adaptiveness, and real-time responsiveness required for efficient crowd delivery and sustainable $\mathcal{SCP}$ outcomes.
\subsection{Our Contributions}
To address the aforementioned limitations, this work proposes a novel \textbf{N}eighbor \textbf{e}mbedded Graph Neural Network-based \textbf{C}rowd \textbf{D}elivery Traffic \textbf{M}anagement (\textbf{NeCDM}) model that achieves the objectives of $SMT$ by delivering robust traffic congestion prediction and optimized route planning for $CDR$. The model further incorporates real-time selection of appropriate delivery vehicles, thereby enhancing both congestion mitigation and delivery efficiency. To the best of the authors’ knowledge, NeCDM is the first comprehensive framework that concurrently integrates traffic congestion prediction and intelligent vehicle selection to support crowd delivery operations in smart city ($SmCt$) environments. The proposed architecture comprises two primary functional components: a \textit{Neighbor-embedded Graph Neural Network-based Traffic Congestion Prediction Unit} (TCPu), responsible for forecasting congestion at delivery stations, and a \textit{Traffic Observation and Management Unit} (TOMu), which facilitates real-time decision-making for selecting the most admissible delivery vehicle. Unlike existing methods, NeCDM provides an adaptive, resilient, and prediction-driven mechanism that ensures operational continuity and superior performance under dynamic urban traffic conditions. The major contributions of this study are summarized as follows:
\begin{itemize}
\item \textit{Traffic Congestion Prediction unit (TCPu) for proactive traffic flow prediction}:  TCPu leverages neighbor-embedded GNN reflecting neighbor nodes' interactions and aggregating nodes' self-features with spatial and temporal traffic details to represent $CDR$ across various delivery stations, significantly enhancing long-term traffic congestion prediction in real-time.
\item  \textit{Traffic Observation and Management unit} (TOMu): This unit offers a sophisticated mechanism for adaptive delivery vehicle choice along delivery routes, traffic rerouting based on detected traffic flow levels, and applying countermeasures to ensure operational continuity thus balancing $\mathcal{SCP}$ (distance covered, travel time, carbon emissions, etc.) for $SmCt$.
\item The amalgamation of TCPu and TOMu units ensures continuous traffic monitoring and adaptability to evolving traffic congestion levels, thereby enhancing the resilience of the smart city ecosystem by facilitating the selection of an appropriate delivery vehicle for each $CDR_i$ pertaining to all routes (both inside and outside); this also curbs the number of on-road vehicles. 
\item \textit{Proven Effectiveness in Test Scenarios}: Implementation and testing of the proposed model using an extended real-world dataset demonstrate its effectiveness in terms of $\pounds$, $\pounds_{rmse}$, and computation time, showcasing superior performance compared to existing methodologies in furnishing $SMT$ goals. 
\end{itemize}
\begin{figure}[!ht] 
	\centering
	\includegraphics[width=\linewidth]{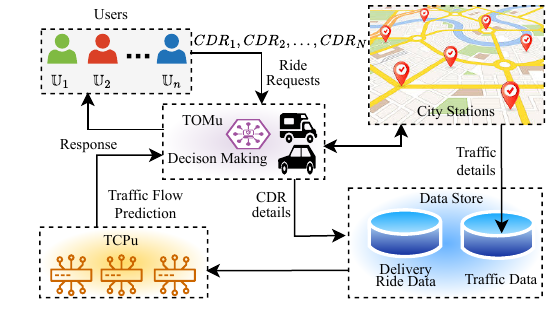}
	\caption{Bird eye view of the proposed NeCDM model.}
	\label{fig:bird}
\end{figure}
A bird's-eye view of the proposed NeCDM model is outlined in Fig. \ref{fig:bird}, where $n$ users $\{\mathbb{U}_1, \mathbb{U}_2, ..., \mathbb{U}_n\}$ raise $N$ crowd delivery ride requests $\{CDR_1, CDR_2, ..., CDR_N\}$ across a $SmCt$, over the TOMu. The spatiotemporal traffic data, along with the $CDR_i$ details, are gathered, quantified, and analyzed to train a significant neighbor-embedded graph neural network-based TCPu periodically. This model simulates the physical real-time traffic management system to provide an accurate prediction of traffic movement across the delivery stations, enabling effective \textit{decision-making} to select a delivery vehicle for a specific $CDR$ at the traffic observation center. 
\subsection{Paper Organization}
Section \ref{sec:rel} discusses the related work. Section \ref{sec:pf} outlines the problem formulation, addressing the highly volatile and evolving traffic congestion challenges in smart city environments, including key assumptions, design goals, and the problem statement. Section \ref{sec:proposed} details the proposed NeCDM model comprising TCP unit for traffic congestion prediction (Section \ref{tcpu}); the TOMu for delivery vehicle selection (Section \ref{dvsu}) and operational design and complexity (Section \ref{sec:opd}), respectively. A comprehensive performance evaluation and baseline comparison are provided in Section \ref{sec:res}. Finally, Section \ref{sec:con} presents the conclusion and the future scope of the proposed work. Table \ref{tab:notation} showcases the symbols with explanatory terms used throughout this article. 
\begin{table}[!htbp]
	\caption{List of Symbols and Notations}
	\label{tab:notation}
	\centering
	\resizebox{0.95\columnwidth}{!}{
		\begin{tabular}{|c||l||c||l|}
			\hline
			\textit{Symbol} & \textit{Notation} & \textit{Symbol} & \textit{Notation} \\
			\hline \hline
			$\mathbb{U}$ & User & $DVeh$ & Delivery vehicle \\
			$D^{St}$ & Delivery station & $SmCt$ & Smart city \\
			$CDR$ & Crowd delivery request & $\mathfrak{D}$ & Distance \\
			$\mathcal{SCP}$ & Smart city parameters & $t_i$ & Time instance \\
			$Ce$ & Carbon emission & $TrT$ & Travel time \\
			$CTr$ & Cost to travel & $S^{St}$ & Source station \\
			$Veh^{L_d}$ & Light delivery vehicle & $Veh^{H_d}$ & Heavy delivery vehicle \\
			$Veh^{fc}$ & Vehicle fuel consumption & $DCov$ & Distance covered \\
			$\mathfrak{X}$ & Latitude & $\Upsilon$ & Longitude \\
			$\ell$ & Layer index & $\sigma$ & Activation function \\
			$\upsilon$ & Station/node & $\epsilon$ & Edge \\
			$\xi$ & Edge set & $\zeta$ & Traffic graph network \\
			$\psi$ & Aggregation function & $\Psi$ & Update function \\
			$\vec{\mathbb{R}}^d$ & Node feature vector & $\vec{\mathbb{R}}^{d'}$ & Route feature vector \\
			$\varrho$ & Traffic flow & $\zeta_e$ & Neighbor-embedded graph \\
			$\acute{\Gamma}$ & Training set & $\ddot{\Gamma}$ & Test set \\
			$\breve{\Gamma}$ & Validation set & $\pounds$ & Loss function \\
			$y_t$ & Actual value & $\hat{y_t}$ & Predicted outcome \\
			$\digamma$ & Prediction function & $\Theta$ & Objective function \\
			$\pounds_{rmse}$ & Root mean square loss & $R$ & Radius of Earth \\
			$SMT$ & Smart mobility and transportation & & \\
			\hline \hline
	\end{tabular}}
\end{table}
\section{Related Work}\label{sec:rel}
Traffic forecasting and congestion management have been extensively studied from multiple perspectives, including hybrid spatio-temporal deep learning models, graph-based and multi-graph neural architectures, and system-level resource-aware deployment frameworks. Recent years have witnessed rapid advances in deep learning–based traffic flow forecasting within intelligent transportation systems (ITS). Early deep models exploited convolutional and recurrent neural architectures to jointly capture spatial and temporal dependencies. Chen et al. developed the PCNN framework for short-term congestion prediction, which encodes periodic traffic patterns via multi-scale convolutions \cite{PCNN-Chen-8392388}. Building on this direction, Ali et al. proposed hybrid spatio-temporal frameworks that explicitly model multiple temporal components and dynamic spatial correlations using attention and graph modules \cite{Ali4-ALI2021852,Ali5-ALI2022233,Awan-9345698}. These methods demonstrate that attention mechanisms and dynamic graph convolutional layers substantially enhance prediction accuracy compared with static CNN–LSTM baselines. The introduction of graph-based spatio-temporal networks such as STGCN and diffusion-graph models further established the effectiveness of graph convolutions in representing road-network topology and flow diffusion, forming the methodological basis for subsequent GNN-driven forecasting approaches \cite{Ali3-10628098,Ali2-10906322}.
\begin{table*}[!htbp]
	\centering
	\caption{Pandect Summary:  Related Work on Traffic Flow Prediction and Intelligent Transportation}
	\label{tab:literature_review}
		\resizebox{\textwidth}{!}{
			\begin{tabular}{|p{2.0cm}||p{2.8cm}||p{3cm}||p{2.2cm}||p{3cm}||p{3cm}|}
				\hline 
				\textit{Approach} & \textit{Method} & \textit{Focus} & \textit{Dataset} & \textit{Contribution} & \textit{Limitations} \\
				\hline \hline
				Ali et al. \cite{Ali1}, 2025 & Attention-based edge--cloud predictor & Energy-efficient resource allocation; urban traffic prediction & IoV traffic datasets & Attention-based predictor with energy-aware scheduling & Simulation-level validation only; lacks multi-region generalization \\
				\hline
				Ali et al. \cite{Ali2-10906322}, 2025 & Attention-driven spatio-temporal deep hybrid (GCN + temporal) & Traffic flow prediction in transportation CPS & METR-LA, PeMS & Combines spatial attention with temporal encoding for high accuracy & Focuses on accuracy; limited deployment and latency analysis \\
				\hline
				Ali et al. \cite{Ali3-10628098}, 2024 & Resource-aware Multi-Graph Neural Network (MGNN) & Multi-access edge computing; spatio-temporal correlations & Urban traffic datasets & Balances prediction accuracy and edge resource usage & Does not consider service migration or mobility patterns \\
				\hline
				Ali et al. \cite{Ali4-ALI2021852}, 2021 & Attention-based dynamic spatio-temporal network & Dynamic spatial correlations; attention mechanism & Taxi trajectory data & Captures dynamic dependencies between urban regions & Static graph structure limits adaptability \\
				\hline
				Ali et al. \cite{Ali5-ALI2022233}, 2022 & Dynamic Spatio-Temporal GCN & Citywide traffic flow prediction & PeMSD7, TaxiNYC & Models temporal trends with dynamic adjacency & High computational cost \\
				\hline
				Awan et al. \cite{Awan-9345698}, 2021 & CNN--LSTM hybrid & Urban flow prediction; dynamic correlations & City traffic data & Combines CNN + LSTM for spatio-temporal modeling & Lacks attention or adaptive spatial weighting \\
				\hline
				Zakarya et al. \cite{Zakarya1-10535446}, 2024 & ApMove: service migration model & MEC-based service migration for connected vehicles & CAV simulations & Reduces latency and energy during migration & Does not incorporate prediction uncertainty or adaptive routing \\
				\hline
				Guo et al. \cite{Guo-7845667}, 2017& Cooperative multi-vehicle routing optimization & Routing; breakdown probability minimization & Simulated road networks & Introduces probabilistic routing to avoid congestion & Offline modeling; lacks real-time learning \\
				\hline
				Zhu et al. \cite{Zhu1-10475356}, 2024 & Graph-Structure Enhanced Pre-trained Language Model & Knowledge graph completion; graph pretraining & Knowledge graph datasets & Improves graph embeddings using pretraining & Not applied to spatio-temporal traffic prediction \\
				\hline
				Saxena et al. \cite{Saxena1-10977971}, 2025 & Multi-depot routing optimization & Smart city routing; congestion and emission reduction & Smart logistics datasets & Integrates congestion and emission objectives into routing & Dependent on external traffic prediction models \\
				\hline
				Hou et al. \cite{Hou1-9805695}, 2022 & Spatio-Temporal GNN & Urban region profiling; graph learning & Urban region datasets & Learns spatial embeddings for analytics and anomaly detection & Focused on profiling, not direct forecasting \\
				\hline
				Chen et al. \cite{PCNN-Chen-8392388}, 2018  & Periodic CNN (PCNN) & Short-term congestion prediction & Traffic loop sensor data & Captures periodic spatio-temporal dependencies & Cannot handle non-periodic anomalies \\
				\hline
				Kouziokas et al. \cite{Kouziokas}, 2021 & Bi-/Uni-directional LSTM & Environmental factor-based forecasting & Traffic + weather datasets & Incorporates environmental inputs into temporal models & Lacks spatial modeling capability \\
				\hline \hline
	\end{tabular}}
\end{table*}
Beyond pure predictive accuracy, a significant research trend focuses on resource-aware and edge/cloud-oriented learning for real-time deployment. Ali et al. designed resource-aware multi-graph neural networks and energy-efficient edge–cloud allocation mechanisms that balance latency, energy, and accuracy for urban traffic prediction \cite{Ali3-10628098,Ali1}. Zakarya et al. introduced ApMove, a migration-aware service-placement technique for connected and autonomous vehicles, minimizing delay and energy consumption in mobile edge computing environments \cite{Zakarya1-10535446}. These studies highlight that prediction and resource management must be co-optimized to sustain QoS in large-scale vehicular networks. Relatedly, Hou et al. employed spatio-temporal GNNs for urban region profiling, demonstrating the transferability of learned spatial representations to higher-level urban analytics \cite{Hou1-9805695}. Parallel advances in pre-trained graph-structured language models \cite{Zhu1-10475356} also inspire knowledge transfer and generalization strategies applicable to spatio-temporal traffic graphs.

Guo et al. minimized road-network breakdown probability through cooperative multi-vehicle routing \cite{Guo-7845667}, while Saxena et al. addressed multi-depot vehicle routing for smart cities under congestion and emission constraints \cite{Saxena1-10977971}. In addition, classical recurrent approaches such as bi- and uni-directional LSTMs incorporating environmental features continue to serve as strong baselines for short-term forecasting \cite{Kouziokas}. Collectively, these works have substantially advanced predictive accuracy, representational richness, and computational efficiency. However, three limitations remain evident: (1) most approaches focus solely on forecasting without closed-loop management, (2) spatial relations are modeled via static or partially dynamic graphs, and (3) deployment or resource optimization is rarely integrated with prediction and control. The proposed NeCDM model addresses these gaps by combining dynamic graph-based congestion prediction (TCPu), optimization-driven traffic management (TOMu), and resource-aware adaptability, thereby providing an end-to-end solution for intelligent vehicular congestion observation and management. A summary of the related work is presented in Table \ref{tab:literature_review}.
\section{Problem Formulation}\label{sec:pf}
A problem configuring the system model, research assumption, challenges, and problem definition with particular design goals is formulated in the following subsections:
\subsection{System Model}
System model comprises five entities, namely: Delivery Stations ($D^{St}$), Users ($\mathbb{U}$), Delivery Vehicles ($DVeh$), Smart City ($SmCt$), and Traffic Observation and Management unit ($TOMu$), that are explained as follows. 
\begin{enumerate}
	\item \textit{Users} ($\mathbb{U}$): An entity that raises a crowd delivery request ($CDR$) to access the best possible delivery route, curtailing $\mathcal{SCP}$. The system model considers $\mathbb{U}$ as a primary entity whose requirements must be implemented for a successful $SmCt$ scenario. 
	\item \textit{Delivery Vehicle} ($DVeh$): An entity, selected for the fulfillment of $CDR$ submitted by $\mathbb{U}$ to the \textit{TOMu} for further services. $DVeh$ can mainly be classified as per fuel type into 'Electric', 'CNG', 'Petrol', and 'Diesel' types with heavy ($Veh^{H_d}$) and low delivery capacities ($Veh^{L_d}$), respectively. 
	\item \textit{Smart City} ($SmCt$): A vehicle traffic movement environment considering essential smart city parameters ($\mathcal{SCP}$), including carbon emission ($Ce$), distance covered ($DCov$), travel time ($TrT$), etc., to fulfill the $SMT$ perspective for each $CDR$.
	\item \textit{Delivery Station} ($D^{St}$): It is an entity that acts as pickup and drop points for a particular $CDR$. The same $D^{St}$ can act as a pickup point for a delivery request and a drop point for some other ride request.
	\item \textit{Traffic Observation and Management unit} ($TOMu$): It is the processing entity in the model responsible for accepting the $CDR$s having single/multiple drop stations requests, from $\mathbb{U}$, and responding to a thorough evaluation of each $CDR_i$.
\end{enumerate}
\subsection{Assumptions and Challenges}
The critical challenges for the NeCDM model to fulfill \{${CDR}_1$, ${CDR}_2$, ..., ${CDR}_N$\} by predicting the traffic congestion and selecting delivery vehicles are described as follows. 
\begin{itemize}	
	\item The existence of \textit{diverse delivery vehicles} $\{DVeh_1, DVeh_2, ..., DVeh_x\} \in DVeh$, poses a challenge when fulfilling crowd delivery requests. These vehicles have different fuel consumption types, which affect their mileage, distance coverage, trip time capacity, and load capability. It implies that each vehicle has unique features that must be considered when selecting the appropriate one for a particular delivery request.
	\item Varying \textit{number of drop station} in different $CDR$ influence the computational complexity of the solution. As a result, it poses critical challenges to the capability of the proposed solution to optimize $\mathcal{SCP}$ within a reasonable time frame \{$t_1$, $t_2$\}.
	\item Highly volatile, constantly changing traffic conditions in $SmCt$ impacts the effectiveness of $CDR$ for generating real-time solution for a reasonable time frame \{$t_1$, $t_2$\}.
	\item To incorporate $\mathcal{SCP}$  such as \textit{carbon emission} to curb air pollution ($Ce$), \textit{distance covered} ($DCov$), \textit{travel time} ($TrT$), and \textit{cost to travel} ($CTr$), collectively into the $CDR$, impose a big challenge on the $CDR$ fulfillment.
	\item Accomplishing each $CDR$ by most appropriate and best-suited single $DVeh$, satisfying all challenges and requirements.
\end{itemize}
The following assumptions are considered to align with the above-mentioned challenges. 
\begin{itemize}
	\item A variety of \textit{vehicle types} based on fuel consumption is limited to $\{Electric, CNG, Petrol, Diesel\}$ only, for a crowd delivery scenario in a smart city $SmCt$.
	\item The \textit{diversity of vehicles} is confined to heavy delivery vehicle ($Veh^{H_d}$) and light delivery vehicle type ($Veh^{L_d}$) such that $Veh^{H_d}$: $\{Petrol, Diesel\}$, and $Veh^{L_d}$: $\{Electric, CNG, Petrol\}$ for any $CDR$.
	\item  The number of \textit{drop points} in the $N$ crowd delivery requests \{${CDR}_1$, ${CDR}_2$, ..., ${CDR}_N$\} are capped to a minimum of single station delivery or up to a maximum of ten drop-stations delivery for practical applications.
	\item The proposed model furnish the priority to the different $\mathcal{SCP}$ while fulfilling the $CDR$s in a $SmCt$. 
	\item The NeCDM model can access the traffic data of $SmCt$ from the \textit{data stores}; ds-RRD and ds-TSTD, to incorporate the impact of volatile traffic situations.
\end{itemize}
\subsection{Problem Statement and Design Goals}
Specifically, the problem is to develop a reasonable solution for the multi-objective constrained $SMT$ problem in a $SmCt$, having numerous delivery stations ($D^{St}$), to meet a $CDR$ along with the selection of appropriate $DVeh$ considering the $SCP$. Given the problem statement, assumptions, and challenges, the proposed model contemplates the following design goals. 
\begin{itemize}
\item \textit{Traffic congestion prediction}: To develop a machine learning-driven model that can determine the long-term traffic before finalization of a delivery route and $DVeh$ for a particular $CDR_i$, considering real-time traffic situations.  
\item \textit{Delivery vehicle selection} ($DVeh$): To develop a decision-making and $CDR$ handling module capable of handling the $CDR$ from different $\mathbb{U}$ and provide an effective response along with the identification of the most appropriate $DVeh$ for the crowd ride request.
\end{itemize}
\section{NeCDM Model}\label{sec:proposed}
\begin{figure*}[!ht] 
	\centering
	\includegraphics[width=\textwidth]{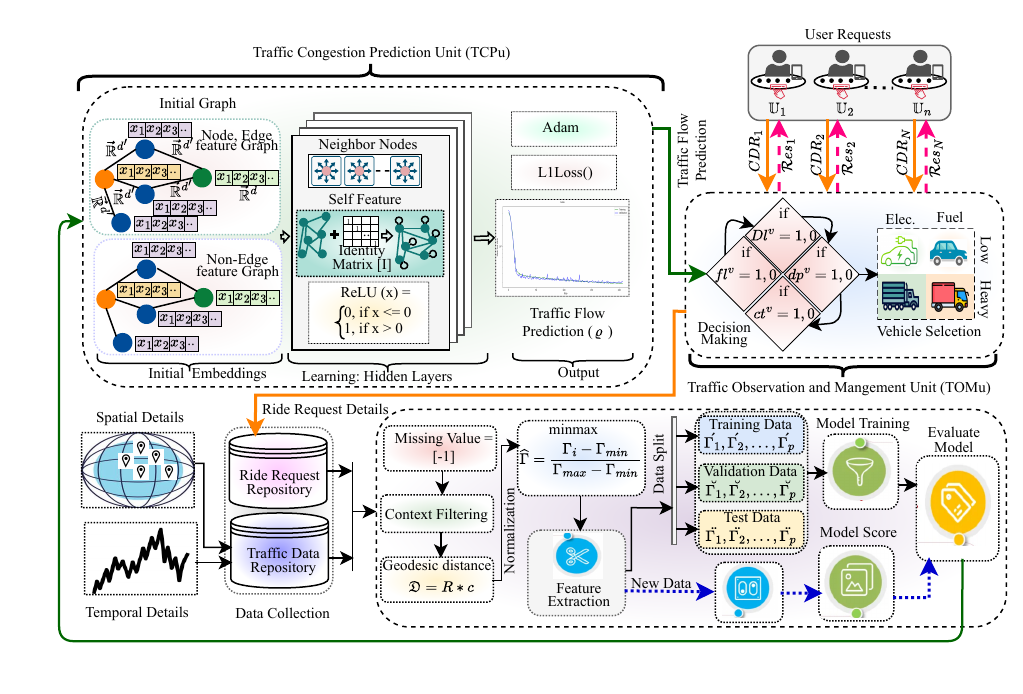}
	\caption{NeCDM model architecture: comprising the TCPu for congestion forecasting across delivery stations, and the TOMu for real-time vehicle selection to optimize crowd delivery, reduce on-road vehicles, and improve smart city parameters.}
	\label{fig:proposed}
\end{figure*}
The proposed model adopts a modular design separating the TCPu and TOMu to enhance interpretability and stability. The TCPu predicts congestion states using spatio-temporal traffic features within a GNN framework, while the TOMu performs decision-level optimization by recommending adaptive routing and control strategies based on these predictions. The TCPu processes real-time vehicular flow data, whereas the TOMu utilizes aggregated traffic indices, predicted congestion maps, and network topology metrics. Integrating these distinct data modalities into a single unified model was found to reduce interpretability and destabilize optimization; hence, the modular design enables sequential learning, where TCPu outputs guide TOMu’s dynamic decision process efficiently.
Consider multiple users $\{\mathbb{U}_1, \mathbb{U}_2, ..., \mathbb{U}_n\}\in \mathbb{U}$ requests for crowd delivery request $CDR$ using \textit{traffic observation and management unit} (TOMu) in a $SmCt$ environment formally defined as follows: 
\begin{definition}
	(\textbf{Smart City}) Let $\textbf{SmCt}$ be a living space that aims to improve residents' quality of life by imparting an Intelligent Transportation System for crowd delivery utilizing data analytics and predictive modeling for automated decision-making to shrink traffic congestion and ensure reliability and availability of transportation. 
\end{definition}
\begin{definition}
	(\textbf{Crowd Delivery Request}) Let $\textbf{CDR}$ be a delivery request $ CDR_i= \{S^{St}_i, (D^{St}_1, D^{St}_2, ..., D^{St}_n)\}$, raised by the user $\mathbb{U}$ where, $S^{St}_i$ represents source station and $D^{St}_n$ specifies $n^{th}$ destination depots of ride request where $n\geq1$, and fulfilled with single $DVeh$. 
\end{definition}
As illustrated in Fig. \ref{fig:proposed}, the $SmCt$ traffic details are managed using two data stores: \textit{Ride Request Details} (ds-RRD) and \textit{Temporal and Spatial Traffic Details} (ds-TSTD). The former maintains records of all delivery requests, while the latter stores spatial data related to the geographical locations of source and delivery stations, along with temporal information (timestamps) capturing historical traffic congestion levels at each station. The NeCDM model deploys a \textit{Traffic Congestion Prediction Unit} (TCPu), which utilizes an advanced neighbor-embedded Graph Neural Network (GNN)-based deep machine learning approach. This unit trains the model using spatiotemporal traffic data in conjunction with ride request data samples and subsequently predicts long-term traffic congestion flow levels ($\varrho$) at each delivery station ($D^{St}$). Additionally, the model incorporates the \textit{Traffic Optimization Module} (TOMu) to select the most appropriate delivery vehicle ($DVeh$) for each customer delivery request ($CDR$). A detailed description of TCPu and TOMu is provided in Sections \ref{tcpu} and \ref{dvsu}, respectively.
\subsection{Neighbor Dynamic Graph Network Architecture}
Let $\zeta(\upsilon,\xi)$ represent the traffic graph, where $\upsilon=\{i\}$ is the set of vertices (delivery stations $D^{St}$), and $\xi=\{\epsilon_{ij}\}$ denotes the set of \textit{un-directed}, \textit{un-weighted}, and \textit{homogeneous} edges (routes) connecting stations $i$ and $j$. The adjacency between any two stations is defined using Eq. (\ref{eq1}):
\begin{equation}  
	\epsilon_{ij}=
	\begin{cases}
		1, & \text{if the stations $i$ and $j$ are connected},\\
		0, & \text{otherwise}.
	\end{cases}
	\label{eq1}
\end{equation}

Each node $i \in \upsilon$ possesses a feature vector $\chi_i^t \in \mathbb{R}^d$ representing its local traffic attributes at time $t$ (e.g., vehicle count, average speed, and congestion ratio), while each edge $\epsilon_{ij} \in \xi$ has a route feature vector $r_{ij}^t \in \mathbb{R}^{d^{\prime}}$ describing inter-station traffic metrics such as travel time, flow rate, and distance. To capture evolving traffic conditions, the adjacency matrix $A^t$ and the corresponding edge weights $w_{ij}^t$ are dynamically updated with time. A continuous weight function encodes the current traffic state using Eq. (\ref{eq1a}):
\begin{equation}
	w_{ij}^t = \exp\!\Big(-\frac{\tau_{ij}^t}{\kappa}\Big),
	\label{eq1a}
\end{equation}
where $\tau_{ij}^t$ is the observed travel time between $i$ and $j$, and $\kappa$ is a scaling constant. This traffic data-based \textit{Graph Neural Network} is capable of learning the structural information using the \textit{neighborhood} of nodes. 
\begin{definition}
	(\textbf{Neighborhood}) Let $\mathbb{N}_i$ of node $i \in \upsilon$ is set of nodes $j \in \upsilon$ connected to i by an edge as $\mathbb{N}_i = \{j: \epsilon_{ij} \in \xi\}$. 
\end{definition}
The message passing for information sharing among nodes at different layers of the network is computed using Eqs. (\ref{eq2}) and (\ref{eq3}), respectively. Furthermore, as the number of delivery stations increases, the number of nodes participating in communication also grows. This results in a higher volume of features to be processed and increased internal communication overhead.
\begin{equation}
	\label{eq2}
	H_{\upsilon}^{(\ell+1)} = \psi  \{\sigma H_{\upsilon}^{(\ell)} \,\forall \, \upsilon  \, \in  \, \mathbb{N}(\upsilon)\} 
\end{equation}
\begin{equation}
	\label{eq3}
	H_{\upsilon}^{(\ell+1)}  = \Psi\{H_{\upsilon}^{(\ell)}, H_{\upsilon}^{(\ell+1)}\}
\end{equation}
Where $\psi$ and $\Psi$ are the aggregation and update functions, respectively, that are applied iteratively at the ${\ell}^th$ layer of the traffic graph network for ten neighbor nodes $\mathbb{N}(\upsilon)$ of all $\upsilon$, and $\sigma$ is the ReLU activation function. The dynamic graph allows the updates in each time window to adapt the adjacency and edge weights ($w_{ij}^t$) according to real-time congestion correlations and travel times. As a result, the TCPu continuously adapts to evolving traffic states, ensuring robust congestion prediction under varying temporal conditions. 
\subsection{Traffic Congestion Prediction}\label{tcpu}
The traffic congestion for $N$ crowd delivery requests:  \{${CDR}_1$, ${CDR}_2$, ..., ${CDR}_N$\} submitted to a \textit{Traffic Observation and Management unit} (TOMu) for effective mobilization of the transportation in a $SmCt$ by $n$ users \{$\mathbb{U}_1, \mathbb{U}_2, ..., \mathbb{U}_n$\} is predicted using the \textit{Traffic Congestion Prediction unit} (TCPu). Each $CDR_i$ includes the details of source station $(S^{St})$ and one or more delivery station $(D^{St})$ details such that \{${CDR}_1: [S^{St}_1, (D^{St}_1, D^{St}_2, ..., D^{St}_p)]$, ${CDR}_2: [S^{St}_2, (D^{St}_1, D^{St}_2, ..., D^{St}_q)]$, ..., ${CDR}_N: [S^{St}_N, (D^{St}_1, D^{St}_2, ..., D^{St}_r)]$\} wherein, $p, q, ..., r \geq 1$. The foremost objective is to fulfill $CDR$s with the least traffic congestion instance. 

Consider $\{CDR_1, CDR_2, ..., CDR_N\}$ for different $S^{St}$ and $D^{St}$ across a $SmCt$ are represented in the form of a \textit{Graph}. The length of the shortest path between two stations on earth surface is computed by utilizing spatial details regarding the latitude ($\mathfrak{X}$) and longitude ($\Upsilon$) details of the source station ($S^{St}$: $\mathfrak{X}_1$, $\Upsilon_1$) and delivery station ($D^{St}$: $\mathfrak{X}_2$, $\Upsilon_2$), respectively by applying \textit{geodesic distance} using Eqs. (\ref{eq4})-(\ref{eq8}). 
\begin{gather}
	\label{eq4}
	\mathfrak{D}_{lon} = \left(\Upsilon_2 - \Upsilon_1\right) \\
	\label{eq5}
	\mathfrak{D}_{lat} = \left(\mathfrak{X}_2 - \mathfrak{X}_1 \right) \\
	\label{eq6}
	a = \left(sin (\frac{\mathfrak{D}_{lat}}{2}) \ast 2 + cos(\mathfrak{X}_1) \ast cos(\mathfrak{X}_2) \ast sin(\frac{\mathfrak{D}_{lon}}{2}) \ast 2 \right) \\
	\label{eq7}
	c = \left(2 \ast a \ast tan \ast 2(\sqrt{a}, \sqrt{1-a}) \right) \\
	\label{eq8}
	\mathfrak{D} = R\times c 
\end{gather}
Where $R$ is the approximate radius of the Earth, it is 6373 km. The distance to travel significantly impacts the $\mathcal{SCP}$ and the choice of $DVeh$. 

All missing values in the dataset are handled by replacing them with $[-1]$. Subsequently, \textit{min--max} normalization is applied to rescale the aggregated $t$ data samples $\{\Gamma_1, \Gamma_2, \ldots, \Gamma_t\}$, which exhibit high variance, to the range $[0, 1]$ using Eq.~(\ref{eq9}). Here, $\Gamma_{min}$ and $\Gamma_{max}$ denote the minimum and maximum values in the input dataset, respectively. The resulting normalized vector is represented as $\widehat{\Gamma}$.
\begin{equation}
	\label{eq9}
	\widehat{\Gamma}=\frac{ \Gamma_i - \Gamma_{min}}{\Gamma_{max}-\Gamma_{min}}
\end{equation} 
The normalized data samples ($\widehat{\Gamma}$) for the $CDR_i^{th}$ are bifurcated into three segments including \textit{training data} \{$\acute{\Gamma_1}$, $\acute{\Gamma_2}$, ..., $\acute{\Gamma_p}$\} $\in \hat{\Gamma}$; \textit{validation samples} \{$\breve{\Gamma_1}$, $\breve{\Gamma_2}$, ..., $\breve{\Gamma_q}$\} $\in \hat{\Gamma}$; and \textit{testing samples} \{$\ddot{\Gamma_1}$, $\ddot{\Gamma_2}$, ..., $\ddot{\Gamma_r}$\} $\in \hat{\Gamma}$, respectively. 

The training dataset $\acute{\Gamma} = \{(\chi_t, y_t)\}_{t=1}^n$, where $\chi_t \in \mathbb{\vec{R}}^d$ denotes the input feature vector and $y_t$ represents the corresponding label (output) vector, is used to train the model. The features of neighboring nodes play a crucial role in traffic prediction. However, TCPu also accounts for the importance of a node’s self-features in addition to the features of its neighbors.

Furthermore, the significant impact of neighboring nodes on the learning process using the training dataset $\acute{\Gamma}$ is captured through a neighbor-embedded graph representation ($\zeta_e$), as computed using Eq.~(\ref{eq10}).

\begin{equation}
	\label{eq10}
	\zeta_e = g\biggl(\chi_t; \mathfrak{D} \biggr)
\end{equation}
Corresponding to $\zeta_e$, output $\hat{y}$ is predicted using the Eq. (\ref{eq11}). 
\begin{equation}
	\label{eq11}
	\varrho_{\hat{y}}= \digamma\biggl(\zeta_e\biggr)= \digamma\bigg(g(\chi_t; \mathfrak{D})\bigg)
\end{equation}
The TCPu ultimately learns the traffic congestion flow $\left(\hat{y} = \digamma\!\left(g(\chi_t; \mathfrak{D})\right)\right)$ by utilizing the nearest-neighbor embedded graph representation–based training dataset $\acute{\Gamma} = \{(\chi_t, y_t)\}_{t=1}^n$. The NeCDM model is trained using \textit{Adam optimization} with \textit{ReLU activation}. The prediction function $\digamma$ is optimized by minimizing the objective function ($\Theta$), as computed using Eq.~(\ref{eq12}).

\begin{equation}
		\label{eq12}
	\begin{split}
		\Theta= \frac{1}{n}\sum_{(\chi, y_t)} \pounds\biggl(y_t, \hat{y}_t\biggr)\\
			= \frac{1}{n}\sum_{(\chi, y_t)} \pounds\biggl(y_t, f(g(\chi_t; \mathfrak{D}))\biggr) 
	\end{split}
\end{equation}
where $\pounds$ is $L_1$-Loss function. It simply measures \textit{mean absolute loss}, i.e., the summation of the absolute difference between the actual value ($y_t$) and the predicted traffic outcome ($\hat{y}_t$) using Eq. (\ref{eq13}). 
\begin{equation}
	\label{eq13}
	\begin{split}
		\pounds= \jmath_{i=1}^n\biggl(y_t, \hat{y}_t\biggr)= \{\jmath_1, \jmath_2, ..., \jmath_n\}^{\dagger} \\
		where \,\,\,\,\,\,\, \jmath_n=|\hat{y}_t-y_t| \\
				=\frac{1}{n}\sum_{i=1}^{n}\biggl|\hat{y}_t-y_t\biggr| \\
	\end{split}
\end{equation}
The \textit{root mean square error} ($\pounds_{rmse}$) (L2-error) to evaluate the loss function more clearly is computed using Eq. (\ref{eq14}). 
\begin{equation}
	\label{eq14}
	\pounds_{rmse}	= \sqrt{\frac{\sum_{i=1}^{n}\biggl(\hat{y}_t-y_t\biggr)^2}{n} }
\end{equation}
Once the TCPu is trained using the prediction function ($\digamma$), it provides the estimated traffic flow level ($\varrho$) for a given $CDR_i$. This value of $\varrho$ is subsequently passed to the TOMu to support effective decision-making for selecting the most appropriate delivery vehicle ($DVeh$), as described in the following section.

\subsection{Delivery Vehicle Selection}\label{dvsu}
The TOMu receives the predicted traffic flow ($\varrho$) corresponding to a $CDR_i$. The TOMu ensures the identification of an appropriate delivery vehicle ($DVeh$) for a particular $CDR$, based on a few delivery parameters like \textit{flow limit}, \textit{distance limit}, the number of \textit{drop points}, and \textit{course of travel}, etc. The mathematical formulation and computation of vehicle selection parameters are described as follows: 
\begin{itemize}
	\item \textit{Flow limit} ($fl^{v}$): The flow limit represents the operational running capacity of a $DVeh$ based on its available fuel or energy. For instance, an electric vehicle with a limited distance coverage ($DCov$) may be suitable only for short-distance deliveries. The flow limit is computed using Eq. (\ref{eq15}), which evaluates whether the vehicle’s speed and operational conditions satisfy admissibility criteria.
\begin{equation}
	\label{eq15}
	f_l^v =
	\mathbf{1}_{\{\varrho \geq x\}} =
	\begin{cases}
		1 	(\text{Admissible}), & \text{if } \varrho \geq x \ (\text{km/h})\\
		0, & \text{otherwise}
	\end{cases}
\end{equation}

	\item \textit{Distance limit} ($Dl^{v}$): The maximum distance required to be traveled to fulfill a particular $CDR$ is coverable by the selected $DVeh$ or not. Every delivery vehicle has a maximum limit on the distance it can travel without requiring any additional service, which is computed using Eq. (\ref{eq16}). 

\begin{equation}
	\label{eq16}
	Dl^{v} = 
	\begin{cases} 
		1 (\text{Reachable}), & \mathfrak{D}_{\varrho} \ge DCov_\text{veh} \\ 
		\frac{\mathfrak{D}_{\varrho}}{DCov_\text{veh}}, & \mathfrak{D}_{\varrho_{l,v}} < DCov_\text{veh}
	\end{cases}
\end{equation}
\item \textit{Number of drops} ($dp^{v}$): A $CDR$ with a large number of drop points requires a $DVeh$ with higher power and payload capacity. For instance, $Veh^{H_d}$ is more suitable in such scenarios. The suitability of a $DVeh$, considering the number of drop stations for a given $CDR_i$, is computed using Eq.~(\ref{eq17}).
	\begin{equation}
		\label{eq17}
		dp^{v}=
		\begin{cases}
			\text{Heavy load vehicle} \,(1), & \text{if\,($dp_{\varrho} \geq x$)} \\
			\text{Low load vehicle} \, (0),  & \text{Otherwise}
		\end{cases}
	\end{equation}
	\item \textit{Course of travel} ($ct^v$): The type of travel route, whether primarily within the $SmCt$ premises or outside it, significantly influences vehicle selection. Travel outside the $SmCt$ typically occurs on highways, which generally require less energy. Conversely, within the $SmCt$, travel involves local roads with higher traffic congestion, necessitating delivery vehicles with greater energy capacity and longer delivery times. The vehicle suitability with respect to the $ct^v$ is computed using Eq. (\ref{eq18}).
	\begin{equation}
		\label{eq18}
		ct^{v}=
		\begin{cases}
			\text{Low energy} \,(1), & \text{if}\,(\text{outside}) \\
			\text{High energy} \, (0),  & \text{Otherwise}
		\end{cases}
	\end{equation}
\end{itemize}	
Finally, all parameters that significantly influence the selection of the most suitable$DVeh$ for a given $CDR_i$ are aggregated into a single feasibility score, as aggregated by Eq. (\ref{eq19}).
\begin{equation}
	\label{eq19}
	DVeh_{CDR_i} = \alpha_1 f_l^v + \alpha_2 D_l^v + \alpha_3 dp^v + \alpha_4 ct^v
\end{equation}

where weights $\alpha_j$ are defined as $\alpha_j \in [0,1]$ with the constraint \[ \sum_{j=1}^{4} \alpha_j = 1,\] allowing flexible prioritization of metrics based on vehicle type, delivery distance, or traffic conditions. 
The proposed NeCDM model performs vehicle traffic congestion prediction and selects the most admissible delivery vehicle with higher efficiency, lower prediction error, and reduced computational time. Furthermore, it accounts for the influence of neighboring nodes in conjunction with a node’s self-features during traffic flow prediction learning, thereby enabling intelligent mobilization and transportation in a $SmCt$ environment.
\subsection{Operational Design and Complexity} \label{sec:opd}
The operational summary for the proposed NeCDM model is described with the help of Algorithm \ref{algo:alg1}. 
\begin{algorithm}[!h]
	\SetAlgoLined
	\caption{NeCDM Model: Operational Summary}	\label{algo:alg1}
	\DontPrintSemicolon
	\KwIn{Two information data stores including \textit{Ride Request Details} (ds-RDD), \textit{Temporal and Spatial Traffic Details} (ds-TSTD)}
	\KwOut{Most appropriate $DVeh_{CDR_i}$}
	
	\tcc{\textbf{CDR-Ride Request:}}  
	Users ($\mathbb{U}$) request $N$ Crowd Delivery Requests ($CDR: [S^{St}_1, (D^{St}_1, D^{St}_2, ..., D^{St}_n)]$)\;
	
	\tcc{\textbf{TGN Generation:}}
	\For{each $CDR_i$  \{${CDR}_1$, ${CDR}_2$, ..., ${CDR}_N$\}}{
		Generate traffic graph networks (TGN)s $\zeta_i$ comprising $S^{St}$  and all  intended $D^{St}$ as nodes\;
		\For{each travel station $\{D^{St}_1, D^{St}_2, ..., D^{St}_n\} \in \zeta_i$}{
			\tcc{\textbf{Congestion Prediction:}}
			Start predicting congestion by NeCDM model using Eqs. (\ref{eq10})-(\ref{eq12})\;
			\tcc{\textbf{Performance Evaluation:}}
			Evaluate the performance of NeCDM model using Eq. (\ref{eq13})\;
			\If{desired performance is achieved}{ 
				Deploy the NeCDM model and estimate traffic flow ($\varrho$) using Eq. (\ref{eq12}) \;
			}
			\Else {Re-train NeCDM Model until desired accuracy is achieved}
			
		}
		\tcc{\textbf{Delivery Vehicle Selection:}}
		Accordingly, using Eqs. (\ref{eq15})-(\ref{eq19}) identify the most appropriate $DVeh_{CDR_i}$ as suitable response for \{${CDR}_1$, ${CDR}_2$, ..., ${CDR}_N$\} from various $\{\mathbb{U}_1, \mathbb{U}_2, ..., \mathbb{U}_n\}$\;
}
	\KwRet{$DVeh_{CDR_i}$}
\end{algorithm}
As outlined in the input step, the NeCDM model utilizes two data stores: \textit{Ride Request Details} (ds-RDD) and \textit{Temporal and Spatial Traffic Details} (ds-TSTD). In Step 1, the model assumes that users employ the NeCDM system to submit crowd delivery requests ($CDR$s), which include source and multiple delivery station details as follows: ($CDR: [S^{St}_1, (D^{St}_1, D^{St}_2, ..., D^{St}_n)]$. Each $n$ user generates $N$ crowd delivery requests, \{${CDR}_1$, ${CDR}_2$, ..., ${CDR}_N$\} where the type of vehicle, such as a heavy-duty vehicle, light-duty vehicle, or specific fuel type, is determined based on the requirements of each particular request. In Step 2, traffic graph networks are created using the source and destination details. In Step 3, for each $i^{th}$ request ($CDR_i$), Steps 4 to 14 are executed. In Step 5, a graph neural network-based algorithm is applied to predict the traffic levels for a specific request. Based on this prediction, the appropriate delivery vehicle is identified for all possible requests.

\textit{Time complexity}: Step 2 iterates the Steps 3-14 for $N$ crowd delivery request raised by $n$ users producing a complexity of $\mathcal{O}(N)$. Step 3 generates traffic graphs for crowd delivery requests with a time complexity of $\mathcal{O}(V+E)$. Step 4 repeats Steps 5-14 for each delivery station $n$ and shows the complexity of $\mathcal{O}(n)$. Further, the intervening Steps 4-13 iterate for $n$ travel stations ($D^{st}$), wherein Step 5 computes objectives cost consumes time-complexity of $\mathcal{O}(n)$ using neighbor embedded graph neural network algorithm having the complexity of $\mathcal{O}(k\ast L\ast \eta^2 \ast \mathbb{\vec{R}}^d \ast \mathbb{\vec{R}}^{d^{\prime}})$, where $k$ is neighbors, $L$ is number of layers, $\eta$ is number of nodes, $\mathbb{\vec{R}}^d$ and $\mathbb{\vec{R}}^{d^{\prime}}$ are the length of node and edge features, respectively. The rest of the steps execute with a time complexity of $ \mathcal{O}(1)$. Therefore, the total time complexity of the NeCDM model is$\mathcal{O}(nkNL(\eta^2+E) \mathbb{\vec{R}}^d \mathbb{\vec{R}}^{d^{\prime}})$.  
\section{Performance Evaluation}\label{sec:res}
\subsection{Experimental Setup}
The experimental work is executed on a server machine assembled with two Intel\textsuperscript{\textregistered} Xeon\textsuperscript{\textregistered} Silver 4114 CPU with a 40-core processor and having 2.20 GHz clock speed in a $SmCt$ environment. The simulation machine is deployed with Ubuntu 16.04, a 64-bit LTS operating system comprising 128 GB of main memory RAM. Enactment of the proposed work is carried out using Anaconda Environment, Jupyter Notebook with implementation in Python 3.9 using \textit{PyTorch} to construct a neighbor-embedded graph network with the configuration. Missing values are handled by inducing with -1; thereafter, \textit{minmax} normalization is applied to bring the dataset values in a range [0,1]. Additionally, in the current simulation scenario for experimental purposes, a $CDR$ is considered to have a maximum of 10 drop-stations, for real-time traffic analysis purposes. All networks are trained using \textit{Adam} optimization, \textit{ReLU} activation, and the parameters as described in Table \ref{tab:hyper} are considered based on extensive evaluation. 
\begin{table}[!htbp]
	\caption{Hyper-parameters configuration}
	\label{tab:hyper}
	\begin{center}
		\resizebox{0.99\columnwidth}{!}{
			\begin{tabular}{|l||c||l||c|}\hline 
				\multicolumn{2}{|c||}{\textit{Pre-processing Parameters}} & \multicolumn{2}{c|}{\textit{Model Training Parameters}} \\ \hline \hline
				\textit{Parameter} & \textit{Value} & \textit{Parameter}  & \textit{Value}   \\ \hline \hline
				Minimum observation  & 1000  & Batch size & 128 \\ 
				Train set size  & 80\% & Learning rate & 0.001 \\ 
				Dropout rate & 0.03 &	weight decay &1$\times10^4$ \\ 
				Test set size &  10\% & Epochs &  100  \\ 
				Validation set size &  10\% & Value per epochs & 4  \\ 
				Normalization &Minmax	&	Number workers &8 \\ 
				Optimization& Adam&	Early stop limit & 20 \\ 
				Activation& ReLU&	Max Neighbor& 10 \\ \hline \hline
		\end{tabular}}
	\end{center}
\end{table}
\subsection{Dataset Details}
The performance of the proposed model is evaluated using the extended trajectory dataset of Beijing City, comprising a total of 70124 samples \cite{datasetzheng2011t-drive}. For each run, the dataset is split randomly into three sub-parts with a ratio 80:10:10 indicating 80\% of the data (56100 samples) is utilized for training ($\acute{\Gamma}$), remaining 20\% data is bifurcated equally in two parts comprising 10\% (7012 samples) for validation ($\breve{\Gamma}$), and 10\% (7012 samples) for testing ($\ddot{\Gamma}$). The spatial data set comprises the location details in the form of GPS coordinates (latitude ($\mathfrak{X}$) and longitude ($\Upsilon$) for the 113 delivery stations across the $SmCt$, along with the unique station number, etc. The temporal data set encompasses time-stamped traffic observations collected hourly for a duration of 02-02-2021 23:00 to 23-03-2023 21:00, across different traffic stations. Further, in three different training: testing: validation scenarios, the validation subset was used exclusively for hyper-parameter tuning and early stopping, while the testing subset was used only once after model selection to assess final performance. This approach ensures that the system is not optimized on the same data used for evaluation, thereby avoiding overfitting. 
\subsection{Results}
\begin{figure}[!htbp]
	\begin{tikzpicture}
		\node[inner sep=0pt] (A) {\includegraphics[width=0.95\columnwidth]{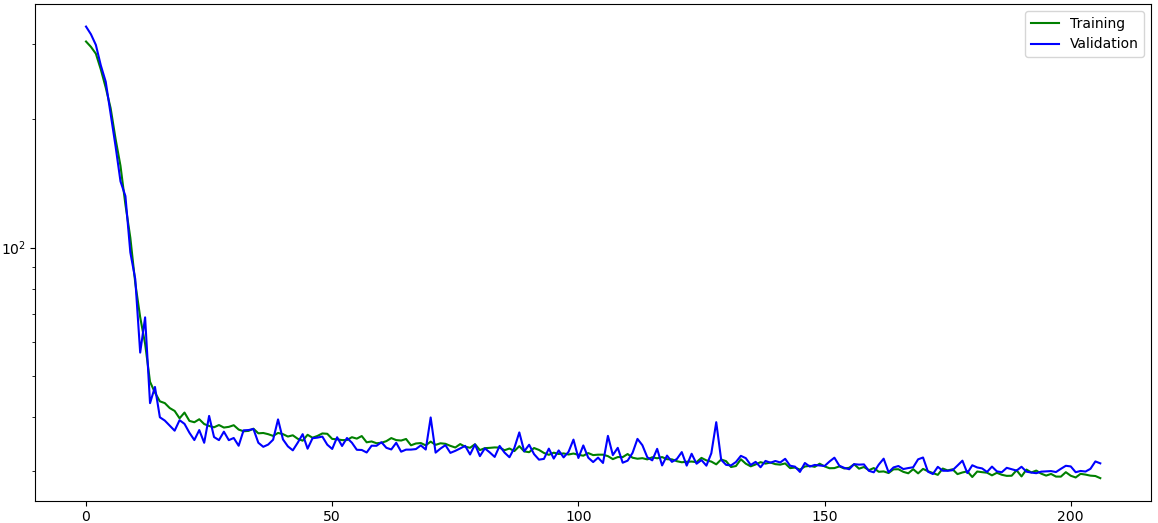}};
		\node[black] (B) at ($(A.south)!-0.05!(A.north)$) {\footnotesize Steps};
		\node[black,rotate=90] (C) at ($(A.west)!-0.03!(A.east)$) {\footnotesize Mean Loss Value ($\pounds$)};
	\end{tikzpicture}
	\caption{Training and validation  L1-loss performance of the proposed NeCDM model across varying iterations.}
	\label{fig:knnloss}
\end{figure}

\textit{Model Optimization}: To predict traffic flow across various delivery stations, the prediction function is trained by minimizing the objective function $\Theta$, resulting in a minimal loss value ($\pounds$). The experimental results illustrating the loss values during the training and validation phases of model development are shown in Fig.~\ref{fig:knnloss}. The horizontal axis represents the training steps [0, 200], while the vertical axis denotes the computed loss value ($\pounds$) on a \textit{logarithmic scale} $(10^2)$. Green line highlights the loss plot during training phase and blue line is for loss plot for validation. It is observed that the loss for the NeCDM model decreases sharply during the initial steps up to 20, followed by a gradual reduction, and finally converges to a minimal value around 200 steps.
\begin{table}[!htbp]
	\centering
		\caption{Traffic prediction loss and execution time performance optimization for a $CDR$}
	\label{tab:epoch}
	\tiny
	\resizebox{0.49\textwidth}{!}{	
		\begin{tabular}{|c||c|c||c||c|c|} 
			\hline
			\multirow{2}{*}{\textit{Epoch}} & \multicolumn{2}{c||}{\textit{Loss ($\pounds$)}} & \textit{Early} & \multicolumn{2}{c|}{\textit{Execution Time}} \\
			\cline{2-3}  \cline{5-6}& $\pounds_{\acute{\Gamma}}$ &$\pounds_{\breve{\Gamma}}$ & \textit{stop} & \textit{Overall (sec)} & \textit{It/s}\\	 \hline \hline
			
			1/100 & 280.9755 &287.1836 & 00/20& 02:00 & 3.19 \\
			10/100 & 60.7781 &53.7435 & 00/20& 01:59 & 3.21 \\
			11/100 & 56.2907 & 52.6622 & 00/20& 01:59 & 3.22 \\ 
			20/100 & 41.6476 & 47.3816 & 01/20& 02:00 & 3.19 \\ \hline \hline
			21/100 & 42.0842 &47.0381 & 03/20& 01:59 & 3.21 \\
			30/100 & 36.9800 &46.5837 & 02/20& 01:59 & 3.21 \\
			31/100 & 37.0872 & 45.1727 & 05/20& 01:59 & 3.20 \\ 
			40/100 & 35.0571 & 43.7102 & 08/20& 01:55 & 3.22 \\ \hline \hline
			41/100 & 34.3449 & 43.6629 & 10/20& 01:58 & 3.25 \\
			50/100 & 33.6158 & 43.3445 & 00/20& 01:59 & 3.22 \\
			51/100 & 33.3891 & 44.7694 & 01/20& 01:59 & 3.23 \\ 
			60/100 & 32.4498 & 42.5025 & 08/20& 02:00 & 3.19 \\ \hline \hline
			61/100 & 32.7173 & 42.5025 & 16/20& 01:59 & 3.21 \\
			70/100 & 32.4312 & 41.3285 & 00/20& 02:00 & 3.20 \\
			71/100 & 31.9725 & 40.5023& 01/20& 01:58 & 3.19 \\ 
			80/100 & 31.5839 & 39.9537& 00/20& 02:00 & 3.20 \\ \hline \hline
			81/100 & 31.2319 & 38.8502& 00/20& 01:59 & 3.19 \\
			90/100 & 30.5238 & 37.2940& 03/20& 02:00 & 3.23 \\
			91/100 & 30.2459 & 36.9472& 07/20& 01:58 & 3.21 \\ 
			100/100 & 29.5213 & 36.3237 & 19/20& 01:45 & 18.26 \\ \hline \hline
	\end{tabular}}
	
	\footnotesize{$\pounds_{\acute{\Gamma}}$: Training loss; $\pounds_{\breve{\Gamma}}$: Validation loss; $sec$: Second; $It/s$: iteration/second} 
\end{table}

Table \ref{tab:epoch} presents the effectiveness of the optimization technique leveraging a neighbor-embedded graph neural network in predicting traffic flow for a crowd delivery request. The results provide insights into the loss points observed during epochs [0, 100]. The loss falls within the range of [280.9755, 29.5213], indicating that the results obtained are not mere chance happenings. Instead, they present a statistically meaningful association within the experimental data. 

\begin{figure}[!htbp]
		\centering
		\begin{subfigure}[t]{0.45\textwidth}
			\centering
				\begin{tikzpicture}
				\node[inner sep=0pt] (A) {\includegraphics[width=0.99\textwidth]{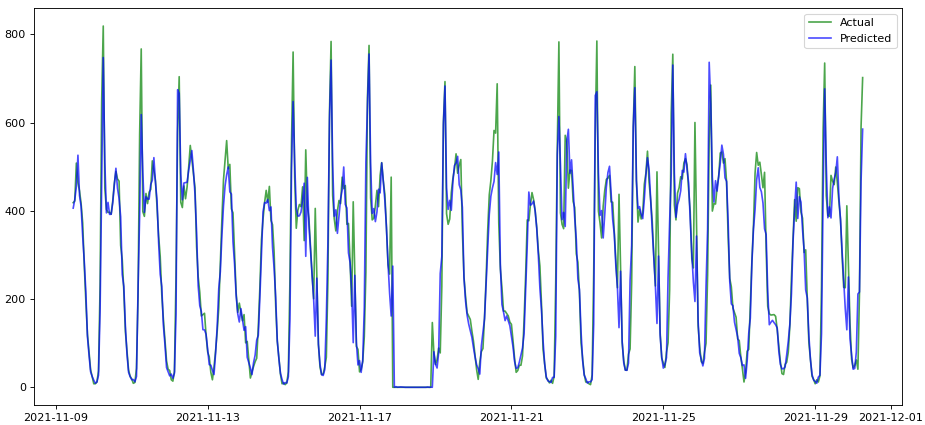}};
				\node[black] (B) at ($(A.south)!-0.05!(A.north)$) {\footnotesize Timeline};
				\node[black,rotate=90] (C) at ($(A.west)!-0.03!(A.east)$) {\footnotesize Traffic flow ($\varrho$)};
			\end{tikzpicture}
			\caption{Traffic flow ($\varrho$) for $D^{st}_1$.}
			\label{fig:knn1}
		\end{subfigure}
				\\
		\begin{subfigure}[t]{0.45\textwidth}
			\centering
			\begin{tikzpicture}
			\node[inner sep=0pt] (A) {\includegraphics[width=0.99\textwidth]{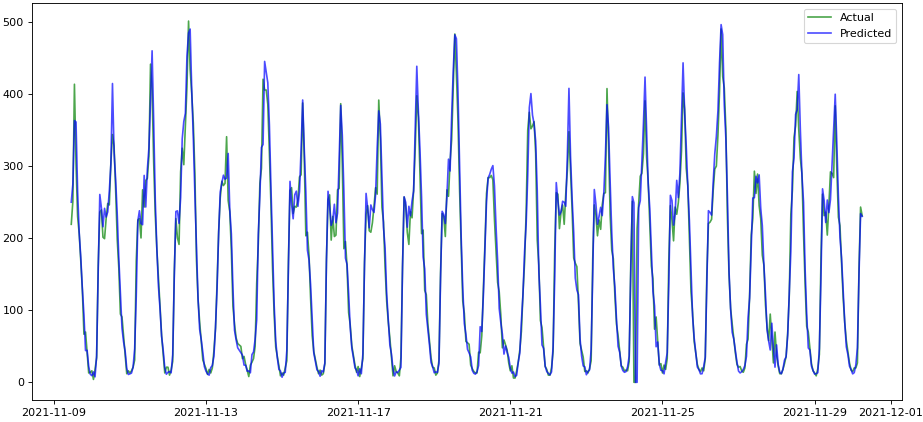}};
			\node[black] (B) at ($(A.south)!-0.05!(A.north)$) {\footnotesize Timeline};
			\node[black,rotate=90] (C) at ($(A.west)!-0.03!(A.east)$) {\footnotesize Traffic flow ($\varrho$)};
		\end{tikzpicture}
		\caption{Traffic flow ($\varrho$) for $D^{st}_2$.}
		\label{fig:knn2}
		\end{subfigure}
			\\
				\begin{subfigure}[t]{0.45\textwidth}
						\centering
						\begin{tikzpicture}
							\node[inner sep=0pt] (A) {\includegraphics[width=0.99\textwidth]{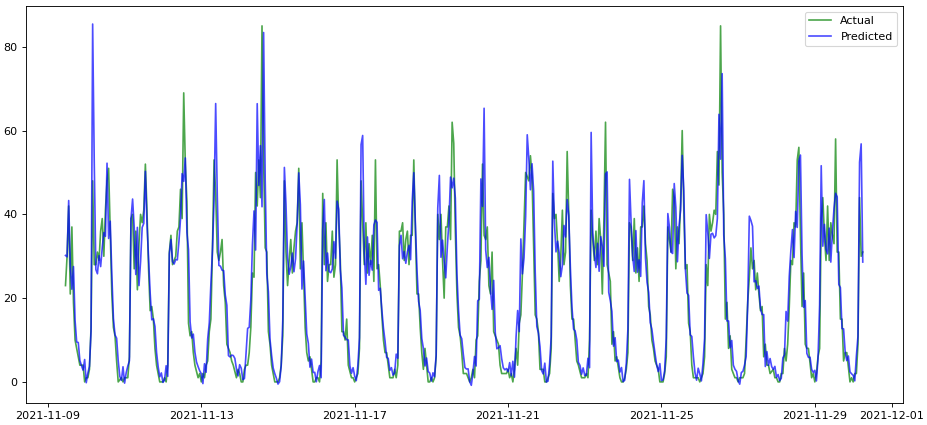}};
							\node[black] (B) at ($(A.south)!-0.05!(A.north)$) {\footnotesize Timeline};
							\node[black,rotate=90] (C) at ($(A.west)!-0.03!(A.east)$) {\footnotesize Traffic flow ($\varrho$)};
						\end{tikzpicture}
					\caption{Traffic flow ($\varrho$) for $D^{st}_3$.}
					\label{fig:knn3}
					\end{subfigure}
		\caption{Traffic flow ($\varrho$) prediction performance across delivery stations over varying timeline.}
		\label{fig:knnpred}
\end{figure}
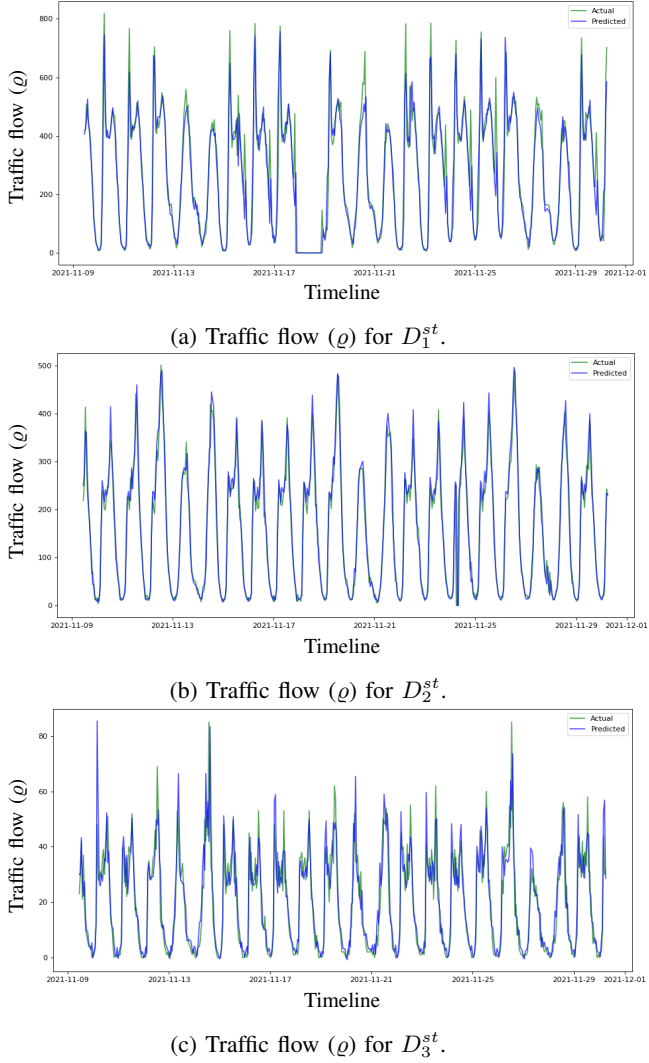

\textit{Traffic Flow Prediction}: The traffic flow ($\varrho$) across delivery stations is illustrated in Fig.~\ref{fig:knnpred}. The timestamp information and traffic flow values are plotted along the X-axis and Y-axis, respectively. The green line represents the actual traffic flow, while the blue line denotes the predicted values. It can be observed that traffic data is unavailable at $D^{St}$ for a certain duration, possibly due to construction work or an accident. Despite this, the proposed model accurately predicts the traffic flow with high precision. This analysis highlights the robustness and effectiveness of the proposed approach in predicting traffic flow across different delivery stations.

\begin{figure}[!htbp]
	\centering
	\begin{subfigure}[t]{0.49\textwidth}
		\includegraphics[width=0.95\columnwidth,height=4.0cm]{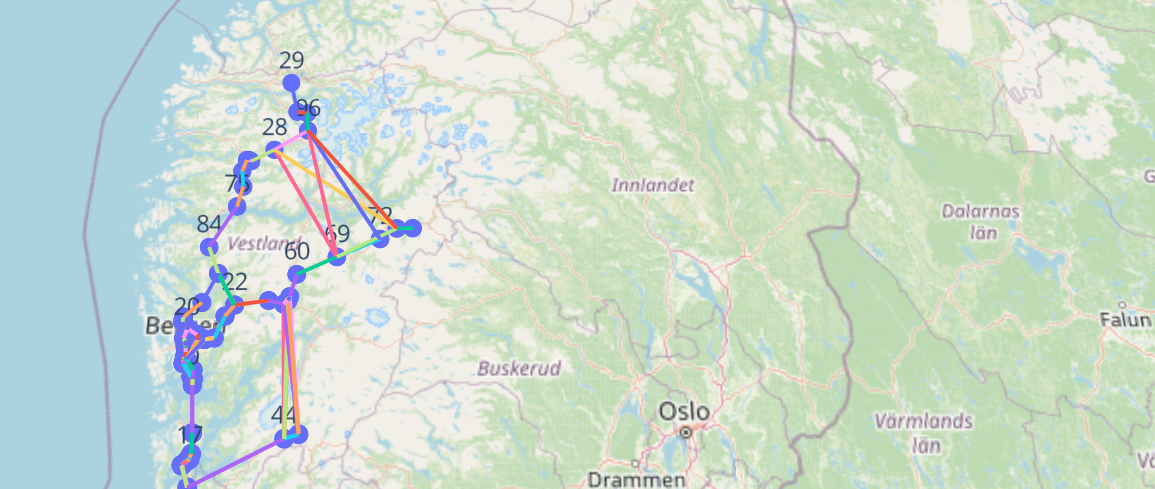}		
		\caption{Zoom-out view.}
		\label{fig:map1}
	\end{subfigure}
	\hfill
	\begin{subfigure}[t]{0.49\textwidth}
		\includegraphics[width=0.95\columnwidth,height=4.0cm]{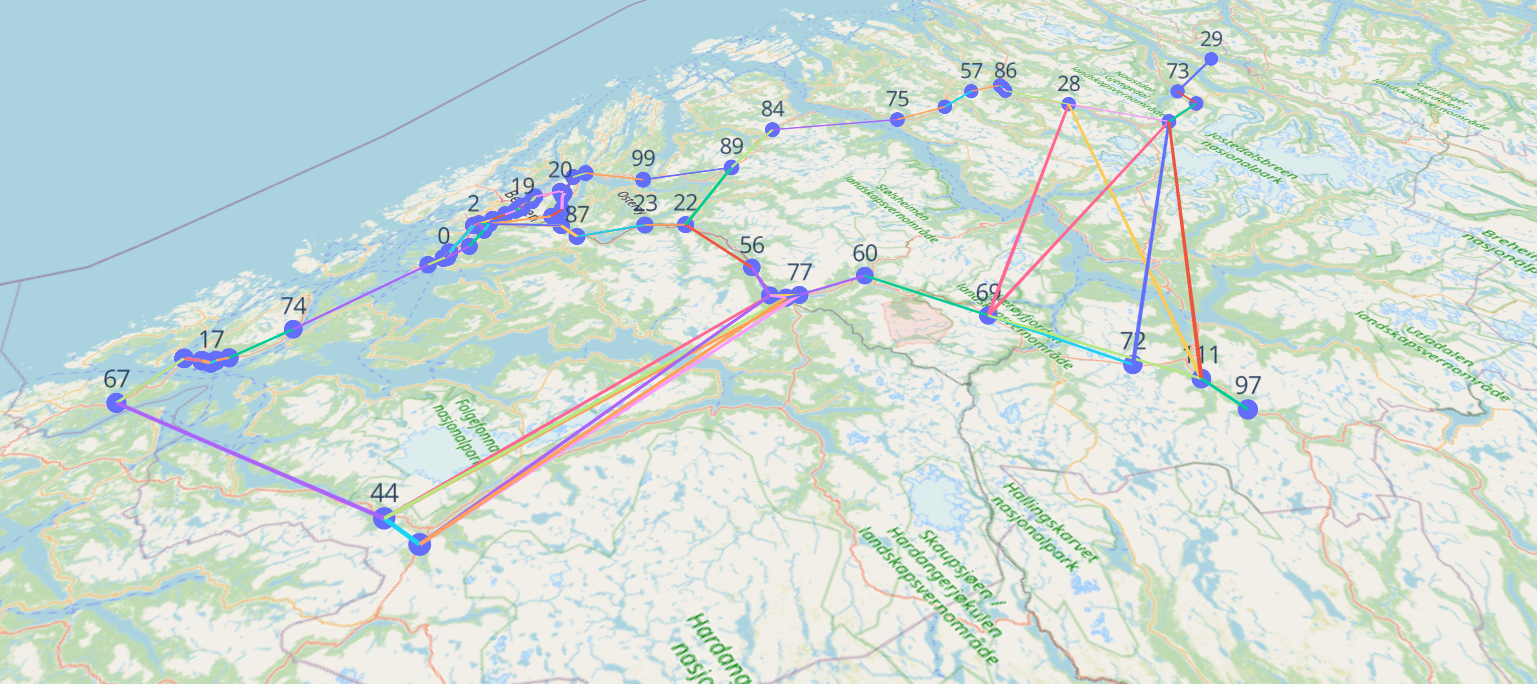}
		\caption{Zoom-in view.}
		\label{fig:map2}
	\end{subfigure}
	\caption{Extended route map depicitng Delivery Station ($D^{St}$) travel route.}
	\label{fig:map}
\end{figure}

\textit{Crowd Delivery Route}: In Fig. \ref{fig:map1} and Fig. \ref{fig:map2}, the achieved travel routes of delivery rides across various $D^{St}$s are shown in both zoomed-out and zoomed-in perspectives, respectively. These are the result of an exhaustive optimization process that implicates the incorporation of neighbor-embedded GNN for traffic flow prediction. 

\textit{Efficiency Analysis}: A detailed efficiency analysis of the proposed NeCDM model under different training scenarios is presented in Table~\ref{tab:train}. In each case, the remaining data are equally divided into validation and testing sets. It is observed that both the L1 ($\pounds$) and L2 ($\pounds_{rmse}$) loss values decrease as the size of the training set increases. This improvement in prediction accuracy can be attributed to the incremental learning capability of the NeCDM model when trained with larger datasets.
\begin{table}[!htbp]
	\caption{Model efficiency analysis with varying traffic data depicting key performance metrics.}\label{tab:train} %
	\centering
	\resizebox{0.98\columnwidth}{!}{
		\tiny
		\begin{tabular}{|c||c|c||c|c|}
			\hline 
			\multirow{2}{*}{\textit{$\acute{\Gamma}:\breve{\Gamma}:\ddot{\Gamma}$}} & \multicolumn{2}{c||}{$Nei_{Ex}$} & \multicolumn{2}{c|}{$Nei_{Em}$}  \\
			\cline{2-5} &  \textbf{\pounds} & $\textbf{\pounds}_{rmse}$ & \textbf{\pounds} &$\textbf{\pounds}_{rmse}$ \\ \hline \hline
			80:10:10 &32.71 &43.22 &29.52&36.32 \\ \hline \hline
			70:15:15 & 32.87&43.96&29.64&37.21 \\ \hline \hline
			50:25:25 &33.53 &44.23&29.70&37.35 \\ \hline \hline
	\end{tabular}}
	\footnotesize{$\acute{\Gamma}$: Training data; $\breve{\Gamma}$: Validation data; $\ddot{\Gamma}$: Test data; $\pounds$: Mean absolute error; $\pounds_{rmse}$: Root mean square error; $Nei_{Ex}$: Neighbor excluded; $Nei_{Em}$: Neighbor embedded} 
\end{table} 
\subsection{Comparison}
The proposed model is compared with two baseline variants: the \textit{Graph Neural Network} (GNN) model and the \textit{Graph Neural Network without Edge Features} (GNN-$Nei_{Ex}$). The GNN model incorporates node features, edge features, and the overall graph structure, whereas the GNN-$Nei_{Ex}$ model follows a similar architecture but excludes edge information. However, neither model explicitly captures the influence of neighboring nodes on the learning process. This limitation is addressed by the NeCDM model, in which the graph is constructed by connecting each node to its nearest neighbors, thereby effectively modeling neighborhood interactions.
\subsubsection{Loss Comparison}
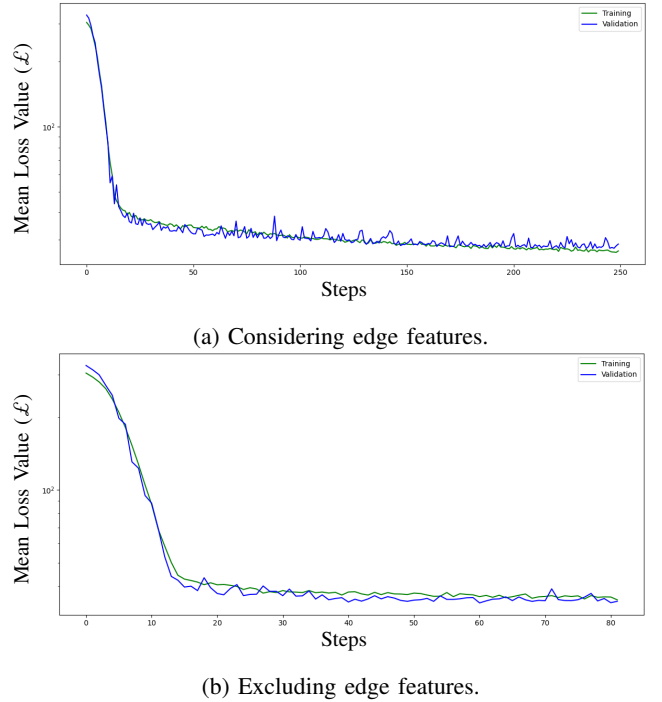
\begin{figure}[!htbp]
	\begin{subfigure}[t]{0.49\textwidth}
	\begin{tikzpicture}
		\node[inner sep=0pt] (A) {\includegraphics[width=0.9\columnwidth]{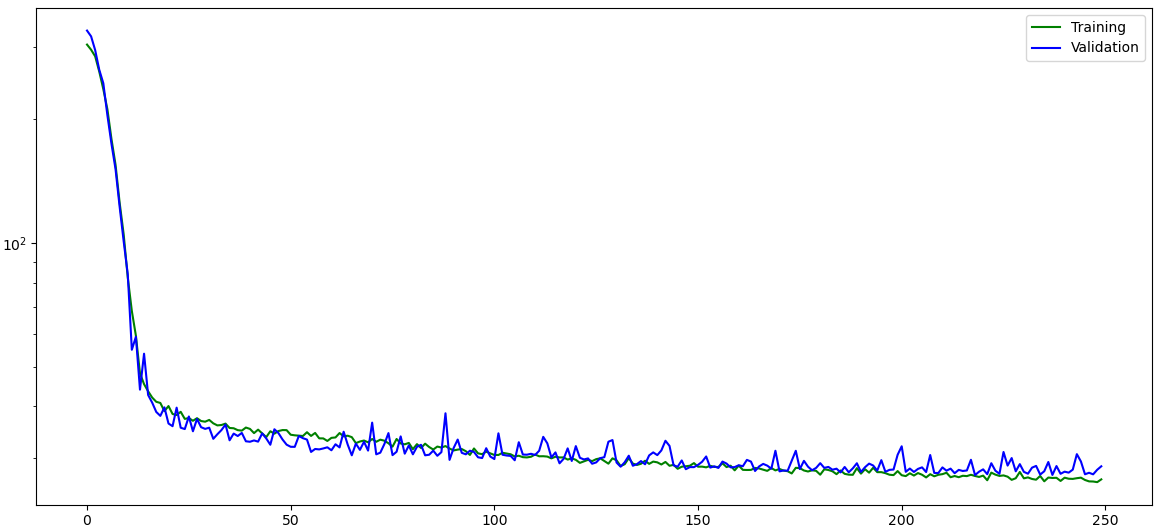}};
		\node[black] (B) at ($(A.south)!-0.05!(A.north)$) {\footnotesize Steps};
		\node[black,rotate=90] (C) at ($(A.west)!-0.03!(A.east)$) {\footnotesize Mean Loss Value ($\pounds$)};
	\end{tikzpicture}
	\caption{Considering edge features.}
	\label{fig:gnnloss}
	\end{subfigure}
	\\ 
	\begin{subfigure}[t]{0.49\textwidth}
	\begin{tikzpicture}
	\node[inner sep=0pt] (A) {\includegraphics[width=0.9\columnwidth]{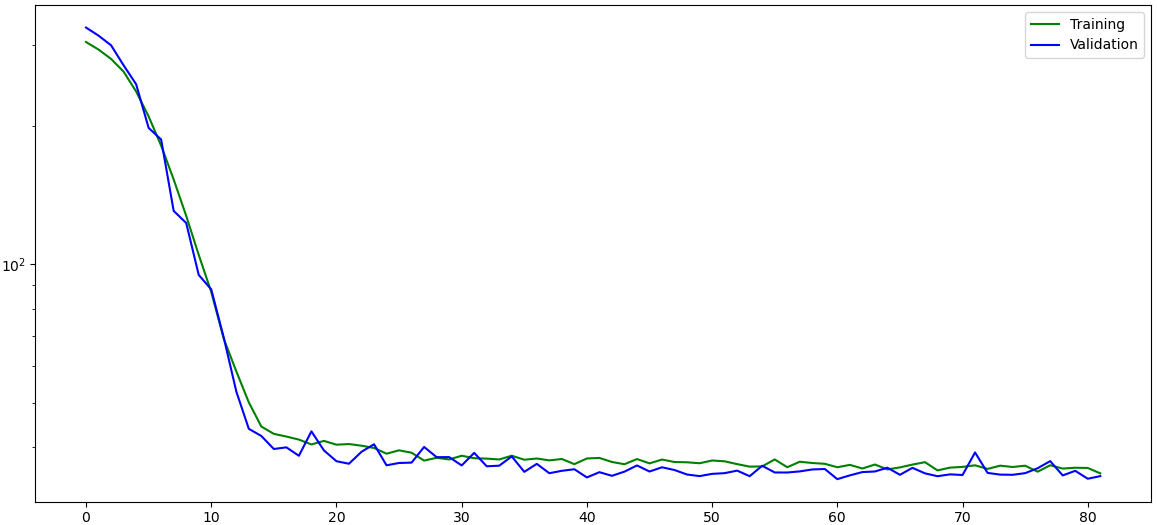}};
	\node[black] (B) at ($(A.south)!-0.05!(A.north)$) {\footnotesize Steps};
	\node[black,rotate=90] (C) at ($(A.west)!-0.03!(A.east)$) {\footnotesize Mean Loss Value ($\pounds$)};
	\end{tikzpicture}
\caption{Excluding edge features.}
\label{fig:gnnneloss}
	\end{subfigure}
		\caption{Traffic prediction model loss analysis for a $CDR$ over varying iteration depicting edge features importance.}
		\label{fig:loss-comp}
\end{figure}
Figs.~\ref{fig:gnnloss} and \ref{fig:gnnneloss} show the loss values for traffic flow prediction using the two baseline models, GNN and GNN-$Nei_{Ex}$, respectively. It can be observed that the training and validation losses decrease progressively with increasing training steps in all models. However, the proposed NeCDM model outperforms both baseline models, achieving greater loss reduction in fewer steps and providing more accurate traffic congestion predictions across $D^{St}$.
\subsubsection{Traffic Flow Comparison}
\begin{figure}[!htbp]
	\centering
	\begin{subfigure}[t]{0.45\textwidth}
		\centering
		\begin{tikzpicture}
			\node[inner sep=0pt] (A) {\includegraphics[width=0.95\textwidth,height=3.3cm]{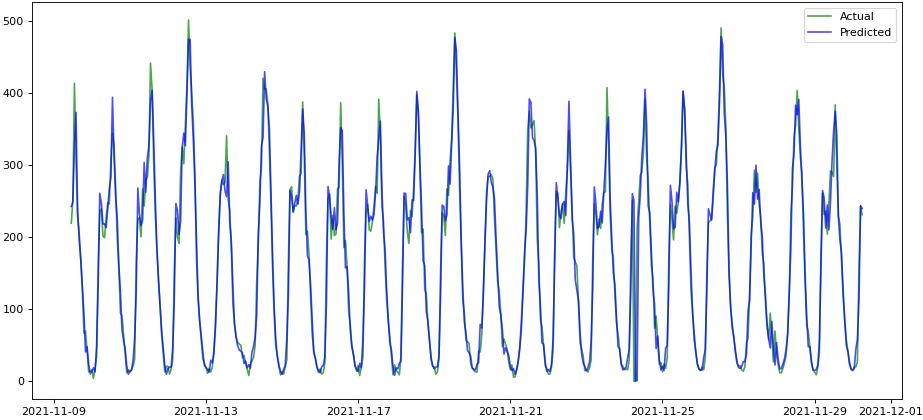}};
			\node[black] (B) at ($(A.south)!-0.05!(A.north)$) {\footnotesize Timeline};
			\node[black,rotate=90] (C) at ($(A.west)!-0.03!(A.east)$) {\footnotesize Traffic Flow ($\varrho$)};
		\end{tikzpicture}
		\caption{Excluding edge feature.}
		\label{fig:gnn_nepred}
	\end{subfigure}
	\\
	\begin{subfigure}[t]{0.45\textwidth}
		\centering
		\begin{tikzpicture}
			\node[inner sep=0pt] (A) {\includegraphics[width=0.95\textwidth,height=3.3cm]{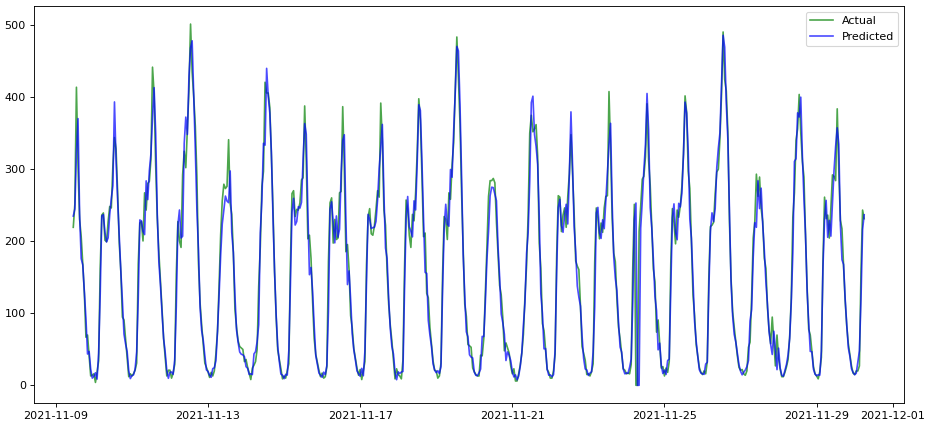}};
			\node[black] (B) at ($(A.south)!-0.05!(A.north)$) {\footnotesize Timeline};
			\node[black,rotate=90] (C) at ($(A.west)!-0.03!(A.east)$) {\footnotesize Traffic Flow ($\varrho$)};
		\end{tikzpicture}
		\caption{Considering all features.}
		\label{fig:gnnpred}
	\end{subfigure}
	\caption{Traffic flow ($\varrho$) prediction performance across $D^{St}$ over varying timeline considering edge features importance.}
	\label{fig:pred}
\end{figure}
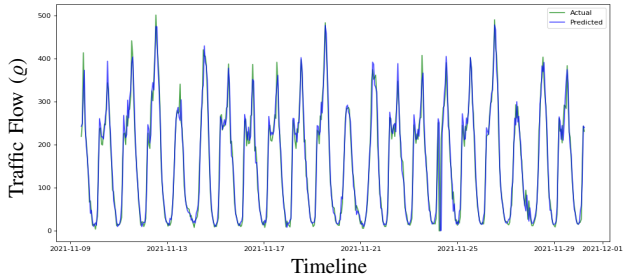
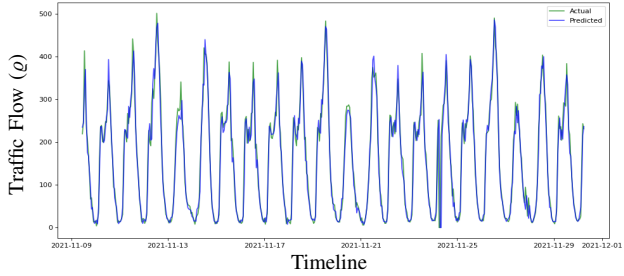

Fig. \ref{fig:gnn_nepred} and Fig. \ref{fig:gnnpred} illustrate the traffic flow at a delivery station ($D^{St}$) computed by the GNN-$Nei_{Ex}$ and GNN baseline models, respectively. It is noticeable that the predicted traffic flow values ($\varrho$) possess some disparities from the actual traffic flow values. The proposed NeCDM model demonstrates better performance by providing more accurate traffic flow predictions, particularly in situations where traffic data is not readily available. The possible reason for this improved performance is that the NeCDM model evaluates the importance of neighbors when learning the traffic patterns, leading to more efficient predictions.
%
\subsubsection{Comprehensive Analysis}
Table \ref{tab:com1} presents the comprehensive analysis comparison of the NeCDM model, by deploying it with comparable existing schemes, including PCNN \cite{PCNN-Chen-8392388}, LSTM \cite{Kouziokas}, GNN \cite{TCSS1-9805695}, and GNN-$Nei_{Ex}$ in terms of computational metrics. It compares the loss and time values of approaches. It can be observed that up to [4.03 to 15.82]\%, [16.66 to 66.63]\%, and [7.64 to 26.78]\% improvement in the proposed approach over the mean absolute loss ($\pounds$), root mean square error ($\pounds_{rmse}$), and overall computation time, respectively. The NeCDM model's capability to consider neighbors' features and their impact on the learning process to predict more realistic traffic flow is the most prominent reason for this outstanding process. 

\begin{table*}[!htbp]
	\caption{Pandect Comparison: NeCDM model v/s state-of-the-art approaches}\label{tab:com1}
	\centering
	\resizebox{0.95\textwidth}{!}{
		\tiny
		\begin{tabular}{|l||l||l||c|c||c||c||l|}
			\hline 
			\multirow{2}{*}{\textit{Approach}} & \multirow{2}{*}{\textit{Method}} & \multirow{2}{*}{\textit{Dataset}} & \multicolumn{2}{c||}{\textit{Loss}} & \textit{Time}&\multirow{2}{*}{\textit{It/s}}  &\multirow{2}{*}{\textit{Complexity}} \\ 
			\cline{4-5} & && $\textit{\pounds}$ &  $\textit{\pounds}_{rmse}$ & \textit{(sec)} & &  \\ \hline \hline
			PCNN \cite{PCNN-Chen-8392388} & Deep convolution network & Vehicle Passage Records & 30.71 &45.75 & 53.93&-&$\mathcal{O}(\Delta tdCN(\eta^2+E)(n^n))$ \\  \hline \hline
			LSTM \cite{Kouziokas} & Long short term memory & Traffic data from Caltrans PeMS in Oakland & 46.73 & 60.41 & 58.36 &-&$\mathcal{O}(\Delta ntd(2h)(\eta^2+E))$ \\  \hline \hline
			
			\multirow{2}{*}{GNN \cite{TCSS1-9805695}} &Graph with node and edge features & Traffic data form Beijing and Hangzhou city & 32.71& 43.22&52.72 &1.63& $\mathcal{O}(n^2NL(\eta^2 +E)  \mathbb{\vec{R}}^d \mathbb{\vec{R}}^{d^{\prime}})$ \\ 
			\cline{2-7}&Graph without edge features & Traffic data form Beijing and Hangzhou city & 31.05& 42.37& 49.55& 1.79&$\mathcal{O}(n^2NL(\eta^2+E)  \mathbb{\vec{R}}^d)$ \\ \hline 
			\textbf{NeCDM} & Neighbor embedded graph & T-Drive trajectory dataset of Beijing city \cite{datasetzheng2011t-drive}   &29.52& 36.32& 46.03 & 3.21& $\mathcal{O}(nkNL(\eta^2+E) \mathbb{\vec{R}}^d \mathbb{\vec{R}}^{d^{\prime}})$ \\  \hline  \hline
	\end{tabular}}
	
	\footnotesize{$\pounds$: loss/Mean absolute error; $\pounds_{rmse}$: Root mean square error; $It/s$: Iteration/second; $sec$:second; PCNN: Deep Convolutional Networks for short-term traffic congestion Prediction; LSTM: Long Short Term Memory; GNN: Graph Neural Network; NeCDM: Proposed Model} 
\end{table*}  
\begin{table*}[htbp]
	\centering
	\caption{Comprehensive Comparison:  NeCDM v/s SoTA approaches depicting model metrics and computational features}
	\label{tab:com2}
\resizebox{\textwidth}{!}{
	\begin{tabular}{|p{1.1cm}||p{2.2cm}||p{2.5cm}||p{1.5cm}||p{1.3cm}||p{1.5cm}||p{1.5cm}||p{1.5cm}||p{3cm}|}
		\hline
		\textit{Approach} & \textit{Spatial Modeling} & \textit{Temporal Modeling} & \textit{Training Loss} & \textit{Testing Loss} & \textit{Time / Epoch (s)} & \textit{Edge/Cloud Awareness} & \textit{Predictive Accuracy} & \textit{Computational Complexity} \\ 
		\hline \hline
		 \cite{Ali1} & Region-level attention; edge nodes & LSTM / temporal fusion & Medium & Medium & Medium & Yes & Medium-High & $O(N^2 + NTd)$ (attention + LSTM) \\
		\hline
		 \cite{Ali2-10906322} & Road-network GCN & LSTM / temporal encoding & Low & Low & High & No & High & $O(EF + NTd)$ (GCN + LSTM) \\
		\hline
		 \cite{Ali3-10628098} & Multi-graph convolution & Temporal CNN / GRU & Low & Low & High & Yes & High & $O(M \cdot EF + NTd)$ (multi-graph + temporal) \\
		\hline
		 \cite{Ali4-ALI2021852} & Dynamic adjacency matrices & LSTM & Low & Low & Medium & No & High & $O(N^2 + NTd)$ (attention + LSTM) \\
		\hline
		 \cite{Ali5-ALI2022233} & GCN layers + adjacency update & LSTM / temporal CNN & Low & Low & High & No & High & $O(EF + NTd)$ (dynamic GCN + LSTM) \\
		\hline
		 \cite{Awan-9345698} & Grid-based CNN & LSTM & Medium & Medium & Medium & No & Medium & $O(W^2D + NTd)$ (CNN + LSTM) \\
		\hline
		 \cite{Zhu1-10475356} & Knowledge graph embeddings & - & - & - & Medium & No & Medium & $O(EF + N^2)$ (graph pretraining) \\
		\hline
		 \cite{Hou1-9805695} & GCN & LSTM / temporal layers & Low & Low & Medium & No & High & $O(EF + NTd)$ (GCN + LSTM) \\
		\hline
		 \cite{PCNN-Chen-8392388} & Grid-based convolution & Temporal CNN & Medium & Medium & Medium & No & Medium & $O(W^2D + T)$ (CNN + temporal encoding) \\
		\hline
		 \cite{Kouziokas} & - & LSTM & Medium & Medium & Low & No & Medium & $O(NTd)$ (LSTM only) \\
		\hline
		 \textbf{NeCDM} & Neighbor embedding & GNN & Low & Low & Low & Yes & High & $\mathcal{O}(nkNL(\eta^2+E) \mathbb{\vec{R}}^d \mathbb{\vec{R}}^{d^{\prime}})$  \\
		 \hline \hline
	\end{tabular}}
	\\[2mm]
	\footnotesize{$N$ = number of nodes/regions, $E$ = number of edges, $F$ = feature dimension, $T$ = temporal sequence length, $d$ = hidden units in LSTM/GRU, $M$ = number of graphs in multi-graph models, $W,D$ = CNN window size and depth, $V$ = number of vehicles, $S$ = number of services.}
\end{table*}

Table~\ref{tab:com2} highlights key advancements in traffic flow prediction for intelligent transportation systems. Dynamic spatio-temporal graph neural networks \cite{Ali2-10906322, Ali3-10628098, Ali5-ALI2022233, Hou1-9805695} excel in predictive accuracy and scalability by capturing complex spatial and temporal traffic patterns. Attention mechanisms \cite{Ali1, Ali4-ALI2021852} enhance the understanding of non-uniform traffic correlations, while edge- and IoT-aware approaches \cite{Ali1, Zakarya1-10535446} support real-time applications in resource-constrained environments. Though CNN-LSTM and LSTM-based methods \cite{Awan-9345698, Kouziokas} are simpler to deploy, they lack adaptability to dynamic traffic. Optimization-based routing methods \cite{Guo-7845667, Saxena1-10977971} are crucial for effective multi-vehicle routing and congestion management, but depend on accurate traffic forecasts. Adopting these advanced methodologies is vital for creating a smarter, more efficient transportation system that meets modern urban mobility needs.
\section{Conclusion}\label{sec:con}
This paper presents a novel NeCDM model that addresses the increasing demand for efficient $SMT$ solutions in $SmCt$ environments. The proposed model is designed to accurately predict traffic congestion across delivery stations and to intelligently select the most suitable delivery vehicle for specific crowd delivery requests. By leveraging neighbor-embedded Graph Neural Networks, the NeCDM model captures dynamic spatial and temporal dependencies to forecast congestion levels, while an integrated decision-making unit facilitates optimal vehicle selection for crowd delivery operations. By effectively reducing the number of active delivery vehicles, the proposed approach helps minimize sustainability-critical parameters ($\mathcal{SCP}$) such as carbon emissions, travel distance, and energy consumption. Extensive performance evaluations demonstrate that NeCDM significantly enhances real-time traffic management, predictive analytics, and system resilience under varying congestion conditions. Future work will focus on integrating quantum computing techniques and scaling the framework to more complex, large-scale urban traffic networks to further enhance its robustness, adaptability, and applicability in next-generation $SMT$ systems.

\bibliographystyle{IEEEtran}
\bibliography{reference}

\begin{IEEEbiography}[{\includegraphics[width=1in,height=1.25in,clip,keepaspectratio]{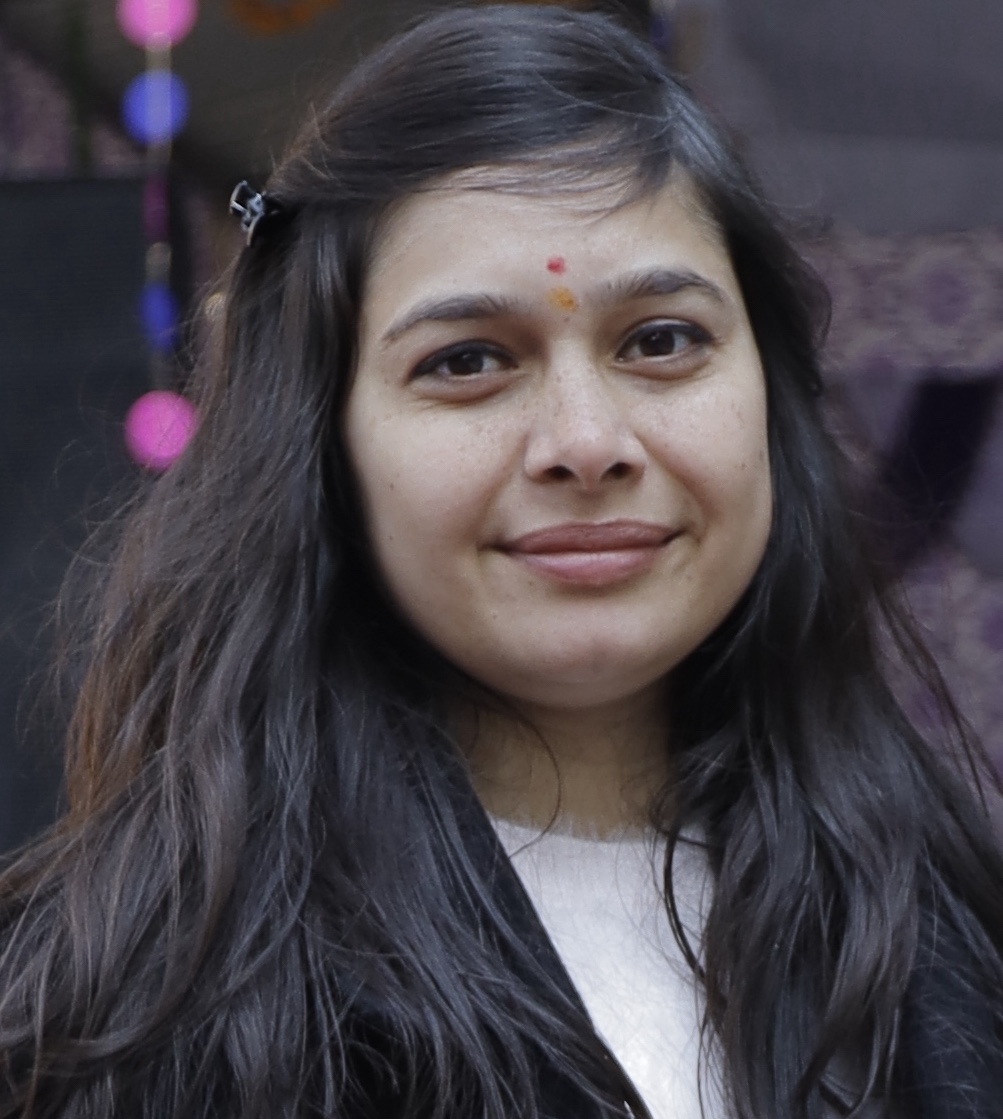}}]{Kishu Gupta} (Member, IEEE) received the Ph.D. degree in computer science from India, a prestigious INSPIRE Fellowship sponsored by the DST, India. She is currently a Postdoctoral Researcher with the National Sun Yat-sen University (NSYSU), Kaohsiung, Taiwan, and associated with VIZJA University, Warsaw, Poland. She has research findings published with top-notch venues, including IEEE TRANSACTIONS ON NEURAL NETWORKS AND LEARNING SYSTEMS, IEEE TRANSACTIONS ON CIRCUITS AND SYSTEMS FOR VIDEO TECHNOLOGY, IEEE TRANSACTIONS ON SYSTEMS, MAN, AND CYBERNETICS, IEEE TRANSACTIONS ON AUTOMATION SCIENCE AND ENGINEERING, IEEE TRANSACTIONS ON INTELLIGENT TRANSPORTATION SYSTEMS, IEEE JOURNAL OF BIOMEDICAL AND HEALTH INFORMATICS, IEEE TRANSACTIONS ON CONSUMER ELECTRONICS, Journal of Network and Computer Applications, Applied Soft Computing, Scientific Reports, Cluster Computing, and Procedia-Computer Science. She was honored with a Gold Medal for her Academic Excellence (Ist Rank in MSc), shortlisted for the Taiwan Comprehensive University System-Young Scholar (TCUS-YS) Award, and was the recipient of the Best Paper Award in RTIP2R-2024 Conference. Her major research interests include data security and privacy, evolutionary optimization, cloud computing, traffic management, federated learning, machine learning, neural networks, and quantum ML
\end{IEEEbiography}
\vspace{11pt}
\begin{IEEEbiography}[{\includegraphics[width=1in,height=1.25in,clip,keepaspectratio]{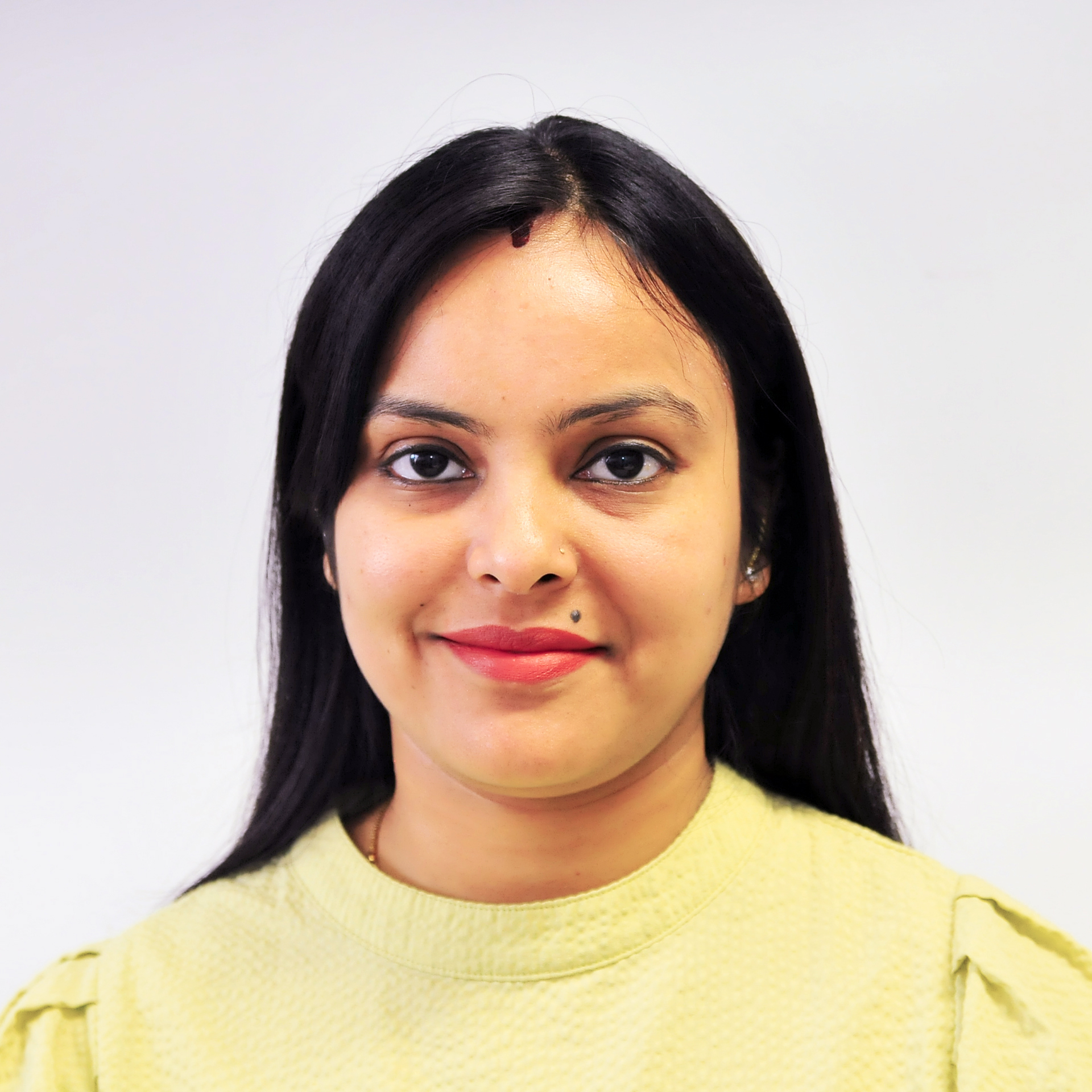}}]{Deepika Saxena} (Senior Member, IEEE) received the Ph.D. degree in computer science from the National Institute of Technology, Kurukshetra, India. She was a Postdoctoral Research Fellow with the Department of Computer Science, Goethe University, Frankfurt, Germany. She is currently an Associate Professor with the Division of Information Systems, The University of Aizu, Japan. She is an Online Lecturer with the VIZJA University, Warsaw, Poland, Europe. Her research interests include neural networks, evolutionary algorithms, resource management and security in cloud computing, internet traffic management, quantum machine learning, data lakes, and dynamic caching management. She was the recipient of the prestigious IEEE TCSC 2023 Outstanding Ph.D. Dissertation Award and EUROSIM 2023 Best Ph.D. Thesis Award, 2022 Best Paper Award for her research article published in IEEE TRANSACTIONS ON CLOUD COMPUTING, and prestigious Japan Society for the Promotion of Science (JSPS) KAKENHI Early Career Young Scientist Research Grant FY2024. 
\end{IEEEbiography}
\vspace{11pt}
\begin{IEEEbiography}[{\includegraphics[width=1in,height=1.25in,clip,keepaspectratio]{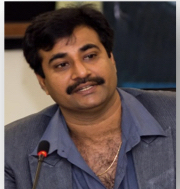}}]{Ashutosh Kumar Singh} (Senior Member, IEEE) received the Ph.D. degree in electronics engineering from the Indian Institute of Technology (BHU) Varanasi, India. He was a Postdoctoral Researcher with the Department of Computer Science, University of Bristol, U.K. He is currently a Professor and the Director of Indian Institute of Information Technology Bhopal, India. Also, he is an Adjunct Professor with the VIZJA University, Warsaw, Poland. He has research and teaching experience in various universities in India, the U.K., and Malaysia. He has authored or coauthored more than 400 research papers in different journals and conferences of high repute. Some of his research findings are published in top cited journals, such as IEEE TRANSACTIONS ON SERVICES COMPUTING, IEEE TRANS- ACTIONS ON COMPUTERS, IEEE TRANSACTIONS ON SYSTEMS, MAN, AND CYBERNETICS, IEEE TRANSACTIONS ON PARALLEL AND DISTRIBUTED SYSTEMS, IEEE TRANSACTIONS ON INDUSTRIAL INFORMATICS, IEEE TRANSACTIONS ON CLOUD COMPUTING, IEEE COMMUNICATIONS LETTERS, IEEE NETWORKING LETTERS, IEEE DESIGN AND TEST, IEEE SYSTEMS JOURNAL, IEEE WIRELESS COMMUNICATIONS LETTERS, IEEE TRANSACTIONS ON NETWORK AND SERVICE MANAGEMENT, IEEE TRANSACTIONS ON GREEN COMMUNICATIONS AND NETWORKING, IET Electronics Letters, Future Generation Computer Systems, Neurocomputing, Information Sciences, and Information Processing Letters. His research interests include the design and testing of digital circuits, data science, cloud computing, machine learning, and security. His research paper, published in IEEE TRANSACTIONS ON CLOUD COMPUTING JOURNAL was honored with the 2022 Best Paper Award by the IEEE Computer Society Publications Board.
\end{IEEEbiography}
\vspace{11pt}
\begin{IEEEbiography}[{\includegraphics[width=1in,height=1.25in,clip,keepaspectratio]{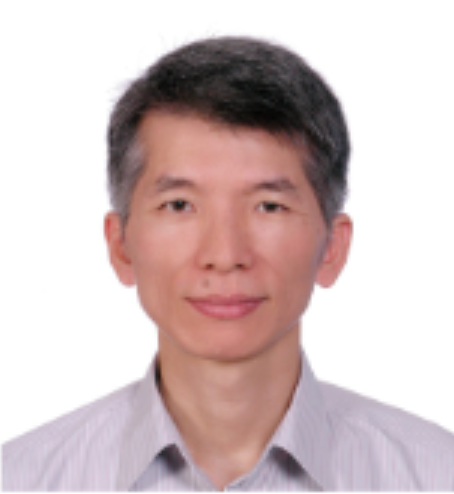}}]{Chung-Nan Lee} (Member, IEEE)received the B.S. and the M.S. degrees in electrical engineering from the National Cheng Kung University, Tainan, Taiwan, in 1980 and 1982, respectively, and the Ph.D. degree in electrical engineering from the University of Washington, Seattle, WA, USA, in 1992. Since 1992, he has been with the National Sun Yat-Sen University, Kaohsiung, Taiwan. He was Head of the Department of Computer Science and Engineering, from 1999 to 2001. He was the President of the Taiwan Association of Cloud Computing, from 2015 to 2017, and the VP for TA of Asia-Pacific Signal and Information Processing Association, from 2019 to 2020. His research interests include multimedia over wireless networks, cloud computing, and IoT. In 2016, he was the recipient of an outstanding engineering Professor Award from the Chinese Institute of Engineers, Taiwan.
\end{IEEEbiography}

\vfill

\end{document}